\documentclass[12pt]{article}

\usepackage{enumerate}
\usepackage{graphicx}
\usepackage{amsmath, amsfonts, amssymb, mathrsfs, rotating, setspace}
\usepackage[utf8]{inputenc}
\usepackage[margin=1in]{geometry}
\usepackage{algorithm2e}
\usepackage{caption}
\usepackage{subcaption}
\usepackage{hyperref}
\usepackage{url}
\usepackage{natbib}
\usepackage{tikz}
\usepackage{soul}
\usepackage{enumitem}

\usetikzlibrary{calc}
\usetikzlibrary{positioning}
\usetikzlibrary{arrows}

\usepackage{amsthm}

\newtheorem{theorem}{Theorem}
\newtheorem{corollary}{Corollary}
\newtheorem{proposition}[theorem]{Proposition}
\renewcommand{\P}{p}
\newcommand{\indep}{\perp \!\!\! \perp}
\newcommand{\blind}{1}

\definecolor{rightcol}{HTML}{FFFF00}
\definecolor{leftcol}{HTML}{0000FF}

\begin{document}

%%%%%%%%%%%%%%%%%%%%%%%%%%%%%%%%%%%%%%%%%%%%%%%%%%%%%%%%%%%%%%%%%%%%%%%%%%%%%%

\def\spacingset#1{\renewcommand{\baselinestretch}%
{#1}\small\normalsize} \spacingset{1}

%%%%%%%%%%%%%%%%%%%%%%%%%%%%%%%%%%%%%%%%%%%%%%%%%%%%%%%%%%%%%%%%%%%%%%%%%%%%%%

\if1\blind
{

  \title{A General Method for Deriving Tight Symbolic Bounds on Causal Effects}
  \author{Michael C. Sachs \and 
  Gustav Jonzon \and
Arvid Sj\"olander \and
Erin E. Gabriel\thanks{The authors report there are no competing interests to declare. MCS and GJ are partially supported by Swedish Research Council grant 2019-00227, EEG by Swedish Research Council grant 2017-01898, and AS by Swedish Research Council grant 2016-01267.} \\
{\footnotesize Department of Medical Epidemiology and Biostatistics, Karolinska Institutet, Stockholm, Sweden} \\
corresponding author: gustav.jonzon@ki.se
}
\maketitle
} \fi

\if0\blind
{
  \bigskip
  \bigskip
  \bigskip
  \begin{center}
    {\LARGE\bf A General Method for Deriving Tight Symbolic Bounds on Causal Effects}
\end{center}
  \medskip
} \fi

\bigskip

\begin{abstract}
%Causal inference involves making a set of assumptions about the nature of things, defining a causal query, and attempting to find estimators of the query based on the distribution of observed variables. When causal queries are not identifiable from the observed data, it may still be possible to derive informative bounds for these quantities in terms of the distribution of observed variables. We develop and describe a general approach for computation of bounds, proving that, if the problem can be stated as a linear program, then the true global extrema result in tight bounds. Building upon previous work in this area, we characterize a class of problems that can always be stated as linear programs; we describe a general algorithm for constructing the linear objective and constraints based on the causal model and query. These problems therefore can be solved using a vertex enumeration algorithm. We develop an R package implementing this algorithm with a user friendly graphical interface using directed acyclic graphs, which only allows for problems within this class to be depicted. We have implemented additional features to help with interpreting and applying the bounds that we illustrate in examples. 
A causal query will commonly not be identifiable from observed data, in which case no estimator of the query can be contrived without further assumptions or measured variables, regardless of the amount or precision of the measurements of observed variables. However, it may still be possible to derive symbolic bounds on the query in terms of the distribution of observed variables. Bounds, numeric or symbolic, can often be more valuable than a statistical estimator derived under implausible assumptions. Symbolic bounds, however, provide a measure of uncertainty and information loss due to the lack of an identifiable estimand even in the absence of data.  We develop and describe a general approach for computation of symbolic bounds and characterize a class of settings in which our method is guaranteed to provide tight valid bounds. This expands the known settings in which tight causal bounds are solutions to linear programs. We also prove that our method can provide valid and possibly informative symbolic bounds that are not guaranteed to be tight in a larger class of problems. We illustrate the use and interpretation of our algorithms in three examples in which we derive novel symbolic bounds.  % with a graphical user interface using directed acyclic graphs and providing additional features to aid in interpretation and application of the bounds, as we illustrate in examples. 
\end{abstract}

\newpage
\spacingset{2}
\vspace{-1cm}

\section{Introduction}
\label{sec:intro}
In many fields of research, a common goal is to determine causal relationships or mechanistic pathways. This investigation is often complicated by common causes of the outcome and either the exposure, or other variables of interest along the causal pathway from the exposure to the outcome, causing confounding. When common causes are unmeasured, the causal effect of interest is usually not identifiable. When the causal effect of interest, which we will refer to as a causal query, cannot be identified, one can derive bounds, i.e., a range of possible values for this quantity in terms of the observed data distribution. 

In general, arbitrarily wide bounds are trivial to derive, but not informative in the sense that they will not provide further insight into the magnitude of the effect. Deriving narrower bounds that are still valid, i.e. containing all possible values of the true causal effect, can be a complicated task, and in particular, deriving tight bounds, i.e., the narrowest possible given all and only explicit assumptions, may be highly non-trivial. An approach to deriving numeric tight bounds in quite general settings is given in \cite{duarte2021automated}. A drawback of the numeric approach, however, is the need for re-computation with each new data set.

Computing bounds symbolically, i.e., as closed form analytic expressions in terms of known observable quantities, rather than numerically, may provide useful information with which to draw conclusions about a study design or form of data collection in the absence of data, in addition to their transparent ease of use in real data once derived. Symbolic tight bounds on a causal query thus, in many ways, provide us with an ideal summary of our effect of interest given our current state of knowledge and/or set of assumptions. 

In 1994, in his PhD dissertation, Alexander Balke gave a method for translating a certain type of causal theory, represented by a directed acyclic graph (DAG), and causal query into a constrained optimization problem \citep{balke1994counterfactual,balke1994probabilistic} in terms of unmeasured response function variables. The causal query is expressed in terms of the distribution of these variables and the DAG gives rise to linear relationships between this distribution and that of the observed variables. In conjunction with standard probabilistic constraints, this yields a bounded constrained optimization problem. If the problem is linear then a vertex enumeration algorithm can be used to find the global extrema of the causal query in terms of the true probability distribution of the observed variables \citep{DantzigLP}. 

%Optimization via vertex enumeration is particular to, and only guaranteed to produce global extrema in, bounded linear optimization problems. 
\citet{balke1994counterfactual} states that the resulting extrema give tight bounds for their causal query in the instrumental variable setting. 
%``Tight'' here means that all values inside the bounds are logically compatible with the true distribution of the observed variables. 
This and related theoretical results have been shown in specific settings that are extensions to the binary instrumental variable problem \citep{ramsahai_causal_2012, bonet2013instrumentality, Heckman01instrumentalvariables}. To the knowledge of the authors, there has been no attempt in the literature to characterize the set of causal problems that are always linear or an approach for determining whether a problem is linear, given its DAG and target query. 

%Balke wrote a program in C\texttt{++} to take a linear programming problem as text file input, perform variable reduction, conversion of equality constraints into inequality constraints, and perform the vertex enumeration algorithm of \citet{mattheiss1973algorithm}. This program has been used by researchers in the field of causal inference with great success \citep{balke1997bounds, cai2008bounds, sjolander2009bounds, sjolander2014bounds} but it is not particularly accessible to other researchers because of the technical challenge of translating the DAG plus causal query into the constrained optimization problem and to determine whether it is linear. Moreover, the program is not optimized and hence does not scale well. Since they only cover a simple instrumental variable setting, it has also not been clear to what extent their techniques extend to more general settings, nor how to apply them to more complex queries. Thus, applications of this approach have been limited to a small number of settings and few attempts to generalize the method to more widely applicable settings and algorithms have been made. 

In this paper, we generalize and extend Balke and Pearl's approach for computation of bounds by characterizing a class of causal problems that always give rise to linear programs and describing a general algorithm for constructing the objective and constraints based on the DAG and query. In \hyperref[discretizationSection]{Section 2}, we introduce the transformation of a causal DAG over categorical variables with unmeasured causal influences into an equivalent one where those influences have been discretized. In \hyperref[classOfDAGsSection]{Section 3} we characterize a set of DAGs that have linear relations between the distributions of their observed variables and unobserved influences, along with an algorithm that extracts those relations from the DAG. \hyperref[classOfQueriesSection]{Section 4} develops notation and requirements for general forms of causal queries that are linear in the distribution of the unmeasured discrete influences, and details an algorithm that constructs such relations from a complex causal query expressed in terms of potential outcomes and observable variables. \hyperref[optimizationSection]{Section 5} then states the final linear program, possible extensions of it and a suitable optimization method. Finally, \hyperref[examplesSection]{Section 6} details a few interesting examples using this method. Proofs of the main propositions are given in \hyperref[appendixA]{Appendix A}. The algorithms described herein are implemented in an R \citep{arr} package called \texttt{causaloptim}, available on the Comprehensive R Archive Network (CRAN), with a user friendly interface. 

%It allows only for problems within this class to be depicted, and is highly optimized. The user inputs the target causal quantity and optionally further linear constraints using standard causal notation.
%Our program then interprets these inputs, automatically converts the problem to the functional model form, and computes the tight bounds using a symbolic linear optimizer. We have implemented a number of additional features to help with interpreting and applying the bounds, for example, the bounds are then functionalized to allow for computation of the bounds with specific values of the parameters, and easy simulation-based testing of the symbolic bounds. 

%We illustrate the steps of the algorithm by using it to derive bounds in a simple example with two confounded variables. Then we apply the method to derive bounds in a novel setting, where there are two instrumental variables that are correlated with each other. 

\section{Discretization} \label{discretizationSection}

%\subsection{Response functional expression of a causal theory}

Let the set of observed variables be denoted $\mathcal{W} = \{W_1, \ldots, W_n\}$, with corresponding vector $\mathbf{W}=(W_1,\dots,W_n)$ and realized values represented by a vector $\mathbf{w}=(w_1, \ldots, w_n)$. We assume that all of these variables are categorical. Each variable of interest $W_i\in\mathcal{W}$ is affected by a set of unmeasured variables $U_{W_i}$ as well as a subset of the remaining variables. Potential outcomes will be denoted using brackets; e.g., $W_1(W_2 = w_2)$, and the probability that the variable $W_1$ would have value $w_1$, if the variable $W_2$ was intervened upon to have value $w_2$ will be denoted as $\P\{W_1(W_2 = w_2) = w_1\}.$ For any given random variable $X$, we let $\nu(X)$ denote its support, i.e., for discrete variables, the set of all values it can take on with positive probability. 

We are assuming the nonparametric structural equation framework, i.e., for each $W_i\in\mathcal{W}$, we assume that there exists a function $F_{W_i}$ such that $w_i$, the value of $W_i$ is given by $w_i = F_{W_i}(\mathbf{pa}_{W_i}, u_{W_i}),$ where $\mathbf{pa}_{W_i}$ denotes the values of variables $\mathbf{Pa}_{W_i}$ in $\mathcal{W}$ that are parents of $W_i$, and $u_{W_i}$ represents the values of $U_{W_i}$, the unmeasured causes of $W_i$. The unmeasured variables $U_{W_i}$ are not assumed independent, unless indicated by the DAG. Since all observed variables of interest in the graph are assumed to be categorical, we can, without loss of generality, recode the assumptions by defining a series of new categorical variables $R_{W_i}$, one for each variable $W_i\in\mathcal{W}$, which specifies how the value of $W_i$ is determined from those of its parents. 
%In this response function variable form of the DAG, if the categorical variable $W_i$ is binary and has $k_i$ parents (excluding the response function variables), then there are $2^{2^{k_i}}$ possible response patterns of $W_i$ with respect to $\mbox{pa}(W_i)$. Thus, we may represent each $R_{W_i}$ as a categorical random variable that takes on $2^{2^{k_i}}$ possible values, one for each response pattern, for $i = 1, \ldots, n$. Let $r_{W_i}$ denote an arbitrary value that the random variable $R_{W_i}$ can take, i.e., each category $r_{W_i}$ occurs with a certain prior probability $\P\{R_{W_i} = r_{W_i}\}$ such that $$\sum_{j=1}^{2^{2^{k_i}}} \P\{R_{W_i} = (r_{W_i})_j\} = 1.$$ Let ${R}$ denote the vector of response function variables $(R_1, \ldots, R_n)$ and ${r} = (r_{W_1}, \ldots, r_{W_n})$ an arbitrary value of the random vector ${R}$. The vector ${r}$ can take on
%\[
%\aleph = \prod_{i = 1} ^ n 2^{2^{k_i}}
%\] 
%possible values.  
%\textbf{Generalization} 
For each $W_i\in\mathcal{W}$, we let $R_{W_i}$ be the variable corresponding to the canonical partition of $\nu(U_{W_i})$ into finite states with respect to the given causal DAG, as stated formally in Proposition \ref{TH1}.
%$=\{r_W:\prod_{V\in\mathbf{pa}(W)}\nu(V)\to\nu(W)\}$ such that for each $r_W\in\nu(R_W)$, there exists a function $f_W$ such that $g_W(\mathbf{pa}_W,U_W,\varepsilon_W)=f_W(\mathbf{pa}(W),r_W)$, so the value $r_W$ of the response function variable $R_W$ (of the measured variable $W$) completely determines the specific dependency of the value $w$ of $W$ on the values of the parents $\mathbf{Pa}_W$ of $W$. 

\begin{proposition}[Canonical partitions] \label{TH1}
\label{responsefunctionvariables}
Let $G$ be a causal DAG, let $\mathcal{W}:=V(G)$ be its vertices and suppose that $\forall W\in\mathcal{W},|\nu(W)|<\infty$ (i.e. each variable is categorical). Let $\mathcal{D} := \nu(\mathbf{Pa}_W)\times\nu(R_W)$ if $\mathbf{Pa}_W$ is nonempty and $\nu(R_W)$ otherwise. Then there exists a categorical variable $R_W$ (so $|\nu(R_w)|<\infty$) and a mapping $f_W:\mathcal{D}\to\nu(W)$ (called the response function of $W$) such that for each value $u_W\in\nu(U_W)$ there exists a unique value $r_W\in\nu(R_W)$ for which $F_W(\cdot,u_W)=f_W(\cdot,r_W)$. %for each value-vector $\mathbf{pa}_W\in\nu(\mathbf{Pa}_W)$ we have $F_W(\mathbf{pa}_W,u_W)=f_W(\mathbf{pa}_W,r_W)$.
\end{proposition}
\noindent For proof, see \hyperref[proofprop1]{Appendix A}.

Regardless of the cardinality of $U_{W_i}$, we have \[
|\nu(R_{W_i})|=|\nu({W_i})|^{|\nu(\mathbf{Pa}_{W_i})|}=|\nu({W_i})|^{\prod_{V\in\mathbf{Pa}_{W_i}}|\nu(V)|}<\infty,\] since all variables in $\mathcal{W}$ are assumed categorical. Let $c_{W_i}:=|\nu({W_i})|$, so $|\nu(R_{W_i})|=c_{W_i}^{\prod_{V\in\mathbf{Pa}_{W_i}}c_V}$, and note that without loss of generality we can assume that $\nu({W_i})=\{0,\dots,c_{W_i}-1\}$ and enumerate $\nu(R_{W_i})$ as $\{0,\dots,c_{W_i}^{\prod_{V\in\mathbf{Pa}_{{W_i}}}c_V}-1\}$. % so \[1=\sum_{r_{W_i}\in \nu(R_{W_i})}p\{R_{W_i}=r_{W_i}\}=\sum_{j=0}^{c_{W_i}^{\prod_{V\in\mathbf{Pa}_{W_i}}c_V}-1}p\{R_{W_i}=j\}.\] 
Let $\mathbf{R}:=(R_{W_1},\dots,R_{W_n})$ and $\aleph:=|\nu(\mathbf{R})|=\prod_{i=1}^{n}|\nu(R_{W_i})|=\prod_{i=1}^{n}c_{W_i}^{\prod_{V\in\mathbf{pa}_{W_i}}c_V}$. 
The joint distribution of $\mathbf{R}$ together with the response functions fully characterize the probabilistic causal model.

%The joint distribution of the response function variables $\P\{{R} = {r}\}$ together with the response functions fully characterize the causal model. To see this, note that given the value ${r}$, all variables $W_i \in \mathcal{W}$ have values that are functionally determined. 
For a given $W_i\in\mathcal{W}$ and fixed $\mathbf{r}\in\nu(\mathbf{R})$, we define a procedure for determining its value $w_i$ by recursively evaluating the corresponding functional expression. Using nested subscripts, we let $W_{i1}, \ldots, W_{ik_i}$ denote the parents of $W_i$ that are in $\mathcal{W}$. Then $w_i$, the value of $W_i$, can be obtained by recursively evaluating 

$$
w_i=g^*_{W_i}(\mathbf{r}) := f_{W_i}(g^*_{W_{i1}}(\mathbf{r}), \ldots, g^*_{W_{ik{_i}}}(\mathbf{r}), r_{W_i}).
$$ 

Any set of observed probabilities can be related to the distribution of response function variables as follows:
$$
\P\{\mathbf{W}=\mathbf{w}\}=\P\{W_1=w_1,\dots,W_n=w_n\}=\sum_{\mathbf{r}\in\nu(\mathbf{R}):\forall i\in\{1,\dots,n\},w_i = g^*_{W_i}(\mathbf{r})}\P\{\mathbf{R}=\mathbf{r}\}.
$$

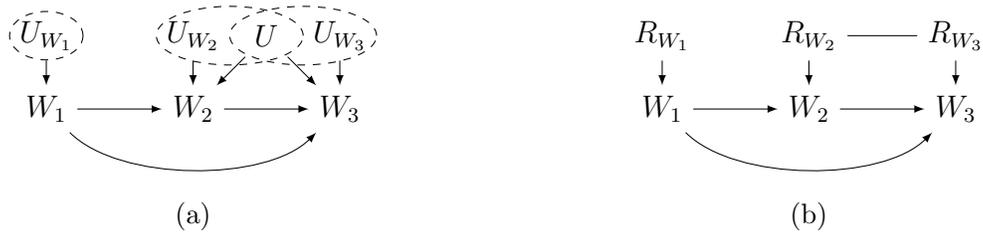
\begin{figure}[h]
\captionsetup[sub]{width=.9\linewidth}
\centering
\resizebox{\linewidth}{!}{
\begin{subfigure}[t]{0.5\textwidth}
\centering
\begin{tikzpicture}
\node (Z) at (-2,0) {$W_1$};
\node (U) at (1,1) {$U$};
\node (X) at (0,0) {$W_2$};
\node (Y) at (2,0) {$W_3$};
\node (uy) at (2,1) {$U_{W_3}$};
\node (uz) at (-2,1) {$U_{W_1}$};
\node (ux) at (0,1) {$U_{W_2}$};
\draw[-latex] (Z) -- (X);
\draw[-latex] (X) -- (Y);
\draw[-latex] (U) -- (X);
\draw[-latex] (U) -- (Y);
\draw[-latex] (uy) -- (Y);
\draw[-latex] (ux) -- (X);
\draw[-latex] (uz) -- (Z);
\draw[-latex] (Z) .. controls (-1,-1) and (1,-1) .. (Y);
\draw[dashed] (1.5,1) ellipse (1cm and .4cm);
\draw[dashed] (.5,1) ellipse (1cm and .4cm);
\draw[dashed] (-2,1) ellipse (.5cm and .33cm);
\end{tikzpicture}
\caption{%Example with unmeasured influences. Dashed ellipses outline unmeasured causal influences. Since those of $W_2$ and $W_3$ both contain $U$, their common cause, their response function variables are dependent. 
\label{a}}
\end{subfigure}

\begin{subfigure}[t]{0.5\textwidth}
\centering
\begin{tikzpicture}

\node (z) at (-2,0) {$W_1$};
\node (rz) at (-2,1) {$R_{W_1}$};
\node (rx) at (0,1) {$R_{W_2}$};
\node (ry) at (2,1) {$R_{W_3}$};
\node (x) at (0,0) {$W_2$};
\node (y) at (2,0) {$W_3$};
\draw[-latex] (x) -- (y);
\draw[-latex] (z) -- (x);
\draw[-latex] (rx) -- (x);
\draw[-latex] (ry) -- (y);
\draw[-latex] (rz) -- (z);
\draw[-latex, -] (rx) -- (ry);
\draw[-latex] (z) .. controls (-1,-1) and (1,-1) .. (y);

\end{tikzpicture}
\caption{%Example with response function variables. If $W_1$, $W_2$ and $W_3$ are binary, the corresponding response function variables $R_1$, $R_2$ and $R_3$ are categorical, taking  $2$, $4$ and $16$ possible values, respectively. 
\label{b}}
\end{subfigure}
}
\caption{Example DAG to illustrate the concepts and notation. In this example, the measured variables are $W_1$, $W_2$, and $W_3$, and the remaining are unmeasured. Since the measured variables are categorical, an equivalent representation of (a) is given in (b), where $R_{W_1}, R_{W_2}, R_{W_2}$ are categorical response function variables. }
\end{figure}

As an example, Figure 1 shows a simple setting with three binary variables of interest. Figure \ref{a} shows the DAG for a model in which variables $W_1$ and $W_2$ both directly affect an outcome $W_3$, with $W_1$ also directly affecting $W_2$. Figure \ref{b} shows the equivalent DAG with response functional variables in place of the original unmeasured variables. The variables that have an unmeasured common cause have response function variables that are dependent, as indicated by the dashed ellipse that outlines the unmeasured causal influences of $W_2$ and $W_3$. Since they both contain $U$, the common cause, their response function variables are dependent as indicated by an undirected edge. We can encode $R_{W_2}$ so the values $0, 1, 2, 3$ of $R_{W_2}$ correspond to the response patterns  $w_2=f_{W_2}(pa_{W_2}\texttt{=} w_1,r_{W_2}\texttt{=} 0):=0, w_2=f_{W_2}(pa_{W_2}\texttt{=} w_1,r_{W_2}\texttt{=} 1):=w_1, w_2=f_{W_2}(pa_{W_2}\texttt{=} w_1,r_{W_2}\texttt{=} 2):=1-w_1, w_2=f_{W_2}(pa_{W_2}\texttt{=} w_1,r_{W_2}\texttt{=} 3):=1$, respectively. We encode $R_{W_1}$ taking values 0 and 1 according to $w_1=f_{W_1}(r_{W_1}\texttt{=}0):=0$ and $w_1=f_{W_1}(r_{W_1}\texttt{=}1):=1$, respectively. Under the model shown in Figure \ref{b}, with e.g. $r_{W_1} = 0$, $r_{W_2} = 1$, $r_{W_3} = 3$, we can evaluate the function to determine $w_2$:
$$
w_2=g^*_{W_2}(\mathbf{r}\texttt{=} (0,1,3)) = f_{W_2}(f_{W_1}(r_{W_1}\texttt{=}0), r_{W_2}\texttt{=}1)) = f_{W_2}(0, 1) = 0.
$$

For $W_3$, we need to enumerate the response patterns for each of the $2^2$ possible combinations of values of $(w_1, w_2)$, i.e., $2^{2^2} = 16$. Then, to evaluate the probability $\P\{W_1 = 1, W_2 = 0, W_3 = 1\}$ in terms of $\mathbf{R}$, we can follow the same procedure as above for all $2^1 \cdot 2^2 \cdot 2^4 = 128$ possible combinations of $\mathbf{r}$, keeping track of the resulting values $\mathbf{w}$. It can be shown that the variable value $\mathbf{w} = (1,0,1)$ is consistent with 16 values of $\mathbf{r}$. Thus the probability of this event is the sum over the set of these 16 values of the probability that $\mathbf{R}$ equals them. See \citet{balke1994counterfactual} or \citet{pearl2009causality}, Chapter 8 for another example and further interpretation.  

%In words and in terms of the algorithm as we implemented it, we evaluate response functions for all possible values of $\mathbf{r}$ given the intervention sets, and find the set of values of $\mathbf{r}$ that yield observations that are consistent with the counterfactual query, ($\mathbf{R}$). Summing those response function variable probabilities gives us the causal query in terms of the joint distribution of the response function variables. The same thing can be done for all observed probabilities to determine a set of equations that relate the observed probabilities to the probabilities that determine the response function variable distribution. These equations combined with the usual probabilistic constraints will be referred to as the constraints. In cases where both the counterfactual query and the constraints are linear, then the bounds can be described in terms of the solution to a linear programming problem. 

%\textbf{Note to self on section 2}: Is the overloaded use of $f$ clear enough?\\
Using this discretization, we can enumerate the relationships between the observable probabilities and the distribution of the response function variables. If those relationships are linear, then they define linear constraints in an optimization problem. Next, we describe a general class of DAGs having linear relationships between their distributions of response function variables and distributions of observable variables.

\section{A Class of Linear DAGs} \label{classOfDAGsSection}

%Next, we describe a general class of DAGs having linear relationships between their distributions of response function variables and distributions of observable variables.
%By ``problem'', we mean the assumptions encoded in a DAG together with a partially observed distribution of its variables (as specified below), the causal query, and optionally additional linear constraints. 
%In causal inference problems, unmeasured common causes (confounding) make the causal effect non-identifiable, which motivates the use of bounds. However, bounds can be improved upon by having a variable that is unconfounded with the outcome of interest \citep{pearl2009causality}. 
To characterize our class of linear problems, the set $\mathcal{W}$ is divided into two subsets $\mathcal{W} = \{\mathcal{W}_\mathcal{L}, \mathcal{W}_\mathcal{R}\}$, where $\mathcal{W}_\mathcal{L}$ may be empty. We assume without loss of generality that the indices of the variables are ordered in such a way that $\mathcal{L} = \{1, \ldots, \gimel\}$ and $\mathcal{R} = \{\gimel+1, \ldots, n\}$, where $\gimel$ may be 0 in which case $\mathcal{L}$ is the empty set. We will denote the corresponding subdivisions of the vectors $\mathbf{W}$ and $\mathbf{R}$ by $(\mathbf{W}_\mathcal{L},\mathbf{W}_\mathcal{R})$ and $(\mathbf{R}_\mathcal{L},\mathbf{R}_\mathcal{R})$, respectively, and likewise for their lowercase value-vector counterparts.  $\mathcal{L}$ and $\mathcal{R}$, connote \emph{left} and \emph{right} sides, where the causal paths flow from left to right. We make this division because in our class of problems, the $\mathcal{L}$-side variables are unconfounded with the $\mathcal{R}$-side variables.

Let $B:=|\nu(\mathbf{W})|=\prod_{i=1}^{n}|\nu(W_i)|=\prod_{i=1}^{n}c_{W_i}$, and let $\{1,\dots,B\}\ni b\mapsto\mathbf{w}_b\in\nu(\mathbf{W})$ be an enumeration of $\nu(\mathbf{W})$ that preserves the ordering of the $\mathcal{L}$-indices before the $\mathcal{R}$-indices such that $\forall b\in\{1,\dots,B\}$, $\mathbf{w}_{b,\mathcal{L}}:=(\mathbf{w}_b)_\mathcal{L}$ and $\mathbf{w}_{b,\mathcal{R}}:=(\mathbf{w}_b)_\mathcal{R}$. Let $\mathbf{p}^*,\mathbf{p}\in[0,1]^B$ be given by $\forall b\in\{1,\dots,B\}$, $p^*_b:=\P\{\mathbf{W}=\mathbf{w}_b\} = \P\{(\mathbf{W}_\mathcal{L}, \mathbf{W}_\mathcal{R}) = (\mathbf{w}_{b,\mathcal{L}}, \mathbf{w}_{b,\mathcal{R}})\}$ and $p_b:=\P\{\mathbf{W}_\mathcal{R}=\mathbf{w}_{b,\mathcal{R}}\mid \mathbf{W}_\mathcal{L}=\mathbf{w}_{b,\mathcal{L}}\}$. 
Thus, the vector $\mathbf{p}^*$ represents the joint distribution of all observed variables and the vector $\mathbf{p}$ contains the observed conditional distribution of all variables in $\mathcal{W}_\mathcal{R}$ given all variables in $\mathcal{W}_\mathcal{L}$. %Note that $\mathbf{p}^*,\mathbf{p}\ge0$, $\sum_{b=1}^Bp^*_b=1$ and each $\sum_{b=K+1}^Bp_b=1$. 
As shown in Proposition \ref{linprobth}, we will only need to observe the components of $\mathbf{p}$.

We will focus on the response function variables of the $\mathcal{R}$-side, and will provide them a dedicated enumeration. Let $$\aleph_\mathcal{R}:=|\nu(\mathbf{R}_\mathcal{R})|=\prod_{i=\gimel+1}^{n}|\nu(R_{W_i})|=\prod_{j=\gimel+1}^{n}c_{W_j}^{\prod_{V\in\mathbf{Pa}_{W_j}}c_V}.$$ Let $\{1,\dots,\aleph_\mathcal{R}\}\ni\gamma\mapsto\mathbf{r}_\gamma\in\nu(\mathbf{R}_\mathcal{R})$ enumerate $\nu(\mathbf{R}_\mathcal{R})$ and $\mathbf{q}\in[0,1]^{\aleph_\mathcal{R}}$ be given by $\forall\gamma\in\{1,\dots,\aleph_\mathcal{R}\}$, $q_\gamma:=\P\{\mathbf{R}_\mathcal{R}=\mathbf{r}_\gamma\}$. 
In particular, the vector $\mathbf{q}$ contains the joint probability distribution of the response function variables $\mathbf{R}_\mathcal{R}$. For $i \in \mathcal{R}$ and a fixed value-vector $\mathbf{w}_\mathcal{L}\in\nu(\mathbf{W}_\mathcal{L})$, we let 
\[
g_{W_i}(\mathbf{w}_\mathcal{L},\mathbf{r}_\gamma):=f_{W_i}(w_{i1}, \ldots, w_{il_i}, g^*_{W_{il_i+1}}(\mathbf{r}_\gamma), \ldots, g^*_{W_{ik_{i}}}(\mathbf{r}_\gamma), r_{W_i})=w_i,
\]
where $w_{i1}, \ldots, w_{il_i}$ are the values of the parents of $W_i$ that are in $\mathcal{W}_\mathcal{L}$, and $W_{il_i+1}, \ldots, W_{ik_{i}}$ are the parents of $W_i$ that are in $\mathcal{W}_\mathcal{R}$.

\begin{proposition} \label{linprobth}
Let $G$ be a causal DAG satisfying the following Conditions:
\begin{enumerate}
\item \label{item:cond1} Any edge that connects two variables $W_\mathcal{L}\in\mathcal{W}_\mathcal{L}$ and $W_\mathcal{R}\in\mathcal{W}_\mathcal{R}$ must be directed from $W_\mathcal{L}$ to $W_\mathcal{R}$. 
%\item There exists an unmeasured variable $U_\mathcal{L}$ such that $U_\mathcal{L}$ is a parent of $W_i$ for all $i \in \mathcal{L}$. That is, all variables in $\mathcal{L}$ share an unmeasured common cause.  
%\item There exists an unmeasured variable $U_\mathcal{R}$ such that $U_\mathcal{R}$ is a parent of $W_i$ for all $i \in \mathcal{R}$. That is, all variables in $\mathcal{R}$ share an unmeasured common cause. 
\item \label{item:cond2} There exists no unmeasured variable $U$ that has children in both $\mathcal{W}_\mathcal{L}$ and $\mathcal{W}_\mathcal{R}$. That is, the variables in $\mathcal{W}_\mathcal{L}$ and $\mathcal{W}_\mathcal{R}$ are not confounded with each other. 
\item \label{item:cond5} There exists an unmeasured variable $U_\mathcal{L}$ such that $U_\mathcal{L}$ is a parent of $W_i$ for all $i \in \mathcal{L}$. That is, all variables in $\mathcal{L}$ share an unmeasured common cause.  
\item \label{item:cond6} There exists an unmeasured variable $U_\mathcal{R}$ such that $U_\mathcal{R}$ is a parent of $W_i$ for all $i \in \mathcal{R}$. That is, all variables in $\mathcal{R}$ share an unmeasured common cause, 
\end{enumerate}
Then there exist matrices $P\in\{0,1\}^{B\times\aleph_\mathcal{R}}$, $P^*\in[0,1]^{B\times\aleph_\mathcal{R}}$ and $\Lambda\in[0,1]^{B\times B}$ such that $\mathbf{p}=P\mathbf{q}$, $\mathbf{p}^*=P^*\mathbf{q}$, $\Lambda$ is diagonal with non-zero diagonal entries, $\Lambda P=P^*$,  $\mathbf{p}^*=\Lambda\mathbf{p}$, and there are no other constraints on the distribution of response function variables that are not redundant with these.
\end{proposition} 

\noindent See \hyperref[proofoflinprobth]{Appendix A} for proof. Conditions \ref{item:cond1} and \ref{item:cond2} ensure that the linear relations are necessary for the distribution to be compatible with the causal model, while the additional conditions \ref{item:cond5} and \ref{item:cond6} ensure that they are also sufficient. Though Proposition \ref{linprobth} guarantees their existence, it may not be trivial to construct these linear relations. Algorithm \ref{probalgo} below details a method for constructing the matrices $P$, $P^*$ and $\Lambda$.
\\

\spacingset{1.25}
\begin{algorithm}[H]
\SetAlgoLined
\KwResult{Systems of linear equations relating $\mathbf{p}^*$ and $\mathbf{p}$ to $\mathbf{q}$}
 Initialize $P$ as a $B\times\aleph_\mathcal{R}$ matrix of $0$s\;
 Initialize ${P^*}$ as a $B\times\aleph_\mathcal{R}$ matrix of $0$s\;
 Initialize $\Lambda$ as a $B\times B$ matrix of $0$s\;
 \For{$b\in1,\dots,B$}{
    \For{$\gamma\in1,\dots,\aleph_\mathcal{R}$}{
        Initialize $\omega$ as an empty vector of length $|\mathcal{R}|\ (=n-\gimel)$\;
        \For{$i\in\mathcal{R}$}{
            Set $\omega_i := g_{W_i}(\mathbf{w}_{b,\mathcal{L}},\mathbf{r}_\gamma)$\;
        }
        \If{$\omega=\mathbf{w}_{b,\mathcal{R}}$}{
            $P_{b,\gamma}:=1$\;
            $\Lambda_{b,b}:=\P\{\mathbf{W}_\mathcal{L}=\mathbf{w}_{b,\mathcal{L}}\}$\;
            $P^*_{b,\gamma}:=\P\{\mathbf{W}_\mathcal{L}=\mathbf{w}_{b,\mathcal{L}}\}$\;
        }
    }
}
\caption{An algorithm to determine a system of linear equations relating $\mathbf{p}$ and $\mathbf{p}^*$ to $\mathbf{q}$.\label{probalgo}}
\end{algorithm}

\spacingset{2}

%We can now summarize our problem as follows. For a given causal DAG satisfying the conditions $(1)-(4)$ given above, an observed conditional distribution vector $\mathbf{p}$ as defined above, a causal query (as specified in the next section) of interest, and optionally additional linear constraints on the vectors $\mathbf{p}$ and $\mathbf{q}$ defined above, find tight bounds on the given query in terms of the components of $\mathbf{p}$. An overview of the algorithm to solve this problem is as follows: (1) For each observed probability conditional on variables in $\mathcal{W}_\mathcal{L}$, convert to linear combination of joint probabilities of the response function variables. (2) Allow for additional linear constraints. (3) Convert the causal query to a linear combination of joint probabilities of the response function variables. (4) Enumerate the vertices of the dual of the linear programming problem. (5) Return bounds in terms of observed probabilities. We describe 1-3 in more detail in turn, and the description of the algorithm serves as a constructive proof that problems in this class correspond to linear programs. 

\section{Functional expressions incorporating interventions} \label{classOfQueriesSection}
%We proceed to determine a set of conditions on a general causal query, under which it may be expressed as a linear combination of response function variables of the $\mathcal{R}$-side. 

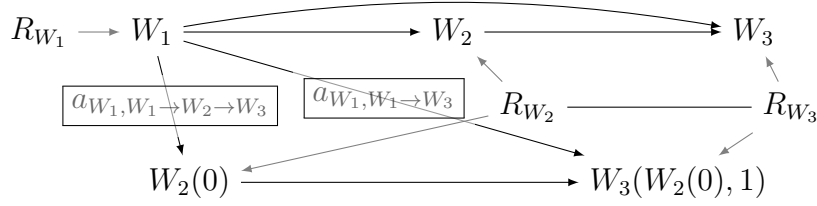
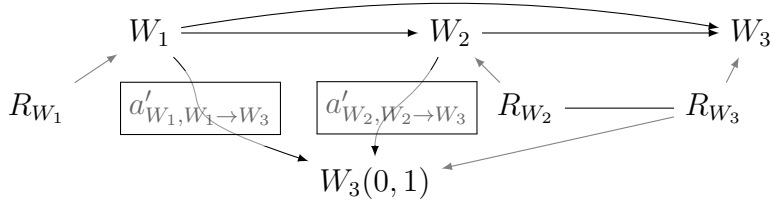
\begin{figure}[h]
\captionsetup[sub]{width=.9\linewidth}
\centering

\begin{subfigure}[t]{0.85\textwidth}
\centering
\begin{tikzpicture}

\node (z) at (-4,0) {$W_1$};
\node (x) at (0,0) {$W_2$};
\node (y) at (4,0) {$W_3$};
\node (y2) at (3,-2) {$W_3(W_2(0), 1)$};
\node (x2) at (-3.5,-2) {$W_2(0)$};
\node (rx) at (1,-1) {$R_{W_2}$};
\node (ry) at (4.5,-1) {$R_{W_3}$};
\node (rz) at (-5.5,0) {$R_{W_1}$};
\draw[-latex] (x) -- (y);
\draw[-latex] (z) -- (x);
\draw[-latex] (z) -- (x2) node [midway, fill=white, fill opacity=.6] {\fbox{$a_{W_1,W_1\rightarrow W_2 \rightarrow W_3}$}};
\draw[-latex] (x2) -- (y2);
\draw[-latex] (z) -- (y2) node [midway, fill=white, fill opacity=.6] {\fbox{$a_{W_1,W_1\rightarrow W_3}$}};
\draw[-latex, thin, gray] (rz) -- (z);
\draw[-latex, thin, gray] (rx) -- (x);
\draw[-latex, thin, gray] (rx) -- (x2);
\draw[-latex, thin, gray] (ry) -- (y);
\draw[-latex, thin, gray] (ry) -- (y2);
\draw[-latex, -] (rx) -- (ry);
\draw[-latex] (z) .. controls (-1,.5) and (1,.5) .. (y);

\end{tikzpicture}
\caption{Extended graph for evaluation of the potential outcome \\ $W_3(W_2(W_1 = 0), W_1 = 1)$. \label{c}}
\end{subfigure}

\begin{subfigure}[t]{0.85\textwidth}
\centering
\begin{tikzpicture}

\node (z) at (-4,0) {$W_1$};
\node (x) at (0,0) {$W_2$};
\node (y) at (4,0) {$W_3$};
\node (y2) at (-1,-2) {$W_3(0, 1)$};
\node (rx) at (1,-1) {$R_{W_2}$};
\node (ry) at (3.5,-1) {$R_{W_3}$};
\node (rz) at (-5.5,-1) {$R_{W_1}$};
\draw[-latex] (x) -- (y);
\draw[-latex] (z) -- (x);
\draw[-latex] (x)  .. controls (-.5,-1) and (-1,-1) .. (y2) node [midway, fill=white, fill opacity=.6] {\fbox{$a'_{W_2,W_2 \rightarrow W_3}$}};
\draw[-latex] (z)  .. controls (-3,-1) and (-4,-1) .. (y2) node [midway, fill=white, fill opacity=.6] {\fbox{$a'_{W_1,W_1 \rightarrow W_3}$}};
\draw[-latex, thin, gray] (rz) -- (z);
\draw[-latex, thin, gray] (rx) -- (x);
\draw[-latex, thin, gray] (ry) -- (y);
\draw[-latex, thin, gray] (ry) -- (y2);
\draw[-latex, -] (rx) -- (ry);
\draw[-latex] (z) .. controls (-1,.5) and (1,.5) .. (y);

\end{tikzpicture}
\caption{Extended graph for evaluation of the potential outcome $W_3(W_2 = 0, W_1 = 1)$. \label{d}}
\end{subfigure}

\caption{Extended DAGs to illustrate that multiple intervention sets are needed to define certain potential outcomes. In these two examples, the variables are binary.}
\end{figure}

In order to determine the values of variables of interest for potential outcomes that incorporate interventions, we must also define a procedure for evaluating a functional expression that allows for variables to be externally forced to certain values. As a first step, we consider extended DAGs, which add additional nodes for potential outcomes of interest as in \citet{balke1994probabilistic}. These are called twin networks in \citet{pearl2009causality}, Chapter 7. Two examples are shown in Figure \ref{c} and \ref{d}. For each potential outcome of interest, nodes are added such that the corresponding factual and potential outcome nodes share the same response function variables. Edges that connect factual nodes to potential outcome nodes are labelled with letters that denote intervention sets indexed by the tail variable of that edge and the path to the head of that edge sequence. These sets define the variables being externally set, the values that they are being set to, and their indices indicate for which edge sequences they apply.

\citet{balke1994counterfactual} considered cases where we externally force a single subset of the variables to some fixed values. This construction suffices for the examples they consider, but not for defining and bounding effects like the natural direct effect of $W_1$ in the graph in Figure \ref{c} whose first term is $\P\{W_3(W_2(W_1 = 0), W_1 = 1) = 1\}$. In that expression, we see that the variable $W_1$, which is a parent of both $W_3$ and $W_2$, is simultaneously being set to 0 and 1, the difference being which child is in question. 
%\citet{sjolander2009bounds} extended the method to work for the natural direct effect, but not more generally than that. 
As another example, the causal query $\P\{W_3(W_2(W_1 = 0)) = 1, W_2(W_1 = 1) = 1\}$ is a joint probability statement, and the two events in question are under different fixed values of $W_1$. Therefore, to be completely general, the variables that one assign to values cannot be a single set; the values that variables are being externally forced to may depend on which children are being considered and also on the term of the probability statement. Thus we define an extended function expression, which ``remembers'' the path of edges taken to get the value that is being determined at each call. 

%\textbf{Note to self}: Does something like this capture what queries look like?:\\
%Queries are of the form $\P\{\xi,\dots,\xi_k\}$, where $\xi,\dots,\xi_k\in\mathcal{Q}=\bigcup_{i=1}^n\mathcal{Q}_i$, and $\mathcal{Q}_1,\dots,\mathcal{Q}_n$ are constructed recursively as follows:\\
%$\forall i\in\{1,\dots,n\},\forall w_i\in\nu(W_i),\quad W_i=w_i\ \in\mathcal{Q}_i$ and \\
%$\forall i\in\{1,\dots,n\}$, if $\mathbf{Pa}(W_i)=\{W_{i1},\dots,W_{ik_i}\}$, then, $\forall\xi_{i1}\in\mathcal{Q}_{i1},\dots,\xi_{ik_i}\in\mathcal{Q}_{ik_i}$, we have $\ W_i(\xi_{i1},\dots,\xi_{ik_i})\ \in\mathcal{Q}_i$.\\

For $i \in \{\gimel+1, \ldots, n\}$, let ${A}_i$ be a matrix that encodes the interventions and variables on which to intervene, with rows indexed by $l$ corresponding to the variables in $\mathcal{W}$ and the columns indexed by $j$ corresponding to all possible paths terminating at $W_i$; the entries in row $l$ are in $\nu(W_l)\cup\{\emptyset\}$. The desired interventions within the causal query then define the entries of $A_i$ which are denoted $a_{lj}$.
%$${A}=\left[\begin{array}{ccc}
%a_{11} & \ldots & a_{1J} \\
%\vdots& \vdots & \vdots\\
%a_{n1} & \ldots & a_{nJ}  \\
%\end{array} \right].$$
In our procedure for evaluating potential outcomes, there is a distinct interventional matrix ${A}_i$ corresponding to each outcome variable $W_i$ used in the causal query. We define the procedure for evaluating the interventional response functional for an outcome variable $W_i$ as

%$$w_i=f_{W_i}\big(w_{i1}\gets f^{{A}}_{W_{i1}}(\mathbf{r}, W_{i1}\rightarrow W_i), \ldots, w_{ik_{i}}\gets f^{{A}}_{W_{ik_{i}}}(\mathbf{r}, W_{ik_{i}} \rightarrow W_i), r_{W_i}\big),$$
$$w_i=h^{A_i}_{W_i}\big(\mathbf{r},W_i\big),$$
where for all $l\in\{1,\dots,n\}$, all $\mathbf{r}\in\nu(\mathbf{R})$ and all strings $j$ representing paths to $W_i$, we define $h^{{A_i}}_{W_i}(\mathbf{r}, j)$ recursively by 

$$
h^{{A_i}}_{W_i}(\mathbf{r},j) \\
:= \begin{cases} a_{lj} & \mbox{ if } a_{lj} \neq \emptyset \\
f_{W_i}(r_{W_i}) & \mbox{ if } a_{lj} = \emptyset \mbox{ and }\mathbf{Pa}_{W_i} = \emptyset \\
%f_{W_i}(w_{i1}\gets f^{{A_i}}_{W_{i1}}(\mathbf{r}, W_{i1} \rightarrow j), \ldots, w_{ik_i}\gets f^{{A_i}}_{W_{ik_i}}\big(\mathbf{r}, W_{ik_{i}} \rightarrow j), r_{W_i}\big) & \mbox{ otherwise, }
f_{W_i}(h^{{A_i}}_{W_{i1}}(\mathbf{r}, W_{i1} \rightarrow j), \ldots,h^{{A_i}}_{W_{ik_i}}\big(\mathbf{r}, W_{ik_{i}} \rightarrow j), r_{W_i}\big) & \mbox{ otherwise, }
\end{cases}
$$
where $k_i:=|\mathbf{Pa}_{W_i}|$ and $\{W_{i1}, \ldots, W_{ik_i}\} := \mathbf{Pa}_{W_i}$, and the notation $i \rightarrow j$ means that $i \rightarrow$ is prepended to $j$. This notation allows us to trace the full path taken from the outcome of interest to the variable being intervened upon. 

For example, considering the DAG in Figure \ref{c} and the causal query $\P\{W_3(W_2(W_1 = 0), W_1 = 1) = 1\}$, we have the interventional matrix 
$${A_3}=\left[\begin{array}{c|cccc}
& W_1\rightarrow W_2 \rightarrow W_3 & W_1 \rightarrow W_3 & W_2 \rightarrow W_3 & W_3\\
\hline
W_1&0&1&\emptyset&\emptyset\\
W_2&\emptyset&\emptyset&\emptyset&\emptyset\\
W_3&\emptyset&\emptyset&\emptyset&\emptyset  \\
\end{array} \right].$$
Thus, evaluating the functional expression $w_3=h^{A_3}_{W_3}(\mathbf{r},W_3)$ results (since $W_3$ is not intervened upon and $\mathbf{Pa}_{W_3}=\{W_1,W_2\}$) in 
$$
w_3 = h^{A_3}_{W_3}(\mathbf{r},W_3) = f_{W_3}(w_1\texttt{=} h^{{A_3}}_{W_1}(\mathbf{r},  W_1\rightarrow W_3), w_2\texttt{=} h^{{A_3}}_{W_2}(\mathbf{r}, W_2 \rightarrow W_3), r_{W_3}).
$$ 
For the first argument of that function call we have $w_1=h^{{A_3}}_{W_1}(\mathbf{r},  W_1\rightarrow W_3) = a_{1,W_1\to W_3} = 1$. Then for the second argument, $a_{2,W_2\to W_3}=\emptyset$ and $\mathbf{Pa}_{W_2}=\{W_1\}$, so we recurse, giving
$$
w_2=h^{{A_3}}_{W_2}(\mathbf{r}, W_2 \rightarrow W_3) = f_{W_2}(w_1\texttt{=} h^{{A_3}}_{W_1}(\mathbf{r}, W_1 \rightarrow W_2 \rightarrow W_3), r_{W_2}).
$$ 
Now, $w_1=h^{{A_3}}_{W_1}(\mathbf{r},  W_1 \rightarrow W_2 \rightarrow W_3) = a_{1,W_1\to W_2\to W_3} = 0$, giving $w_2=f_{W_2}(w_1\texttt{=}0,r_{W_2})$, so we get $w_3=f_{W_3}(w_1\texttt{=}1, w_2\texttt{=} f_{W_2}(w_1\texttt{=}0,r_{W_2}), r_{W_3}).$

For the DAG in Figure \ref{d} and the first part of the causal query $\P\{W_3(W_2=0, W_1 = 1) = 1\}$, we have 
$${A_3}=\left[\begin{array}{c|cccc}
& W_1\rightarrow W_2 \rightarrow W_3 & W_1 \rightarrow W_3 & W_2 \rightarrow W_3 & W_3\\
\hline
W_1&\emptyset&1&\emptyset&\emptyset\\
W_2&\emptyset&\emptyset&0&\emptyset\\
W_3&\emptyset&\emptyset&\emptyset&\emptyset  \\
\end{array} \right].$$

Thus, evaluating the functional expression $w_3=h^{A_3}_{W_3}(\mathbf{r},W_3)$ results in 
$$
w_3 = h^{A_3}_{W_3}(\mathbf{r},W_3) = f_{W_3}(w_1\texttt{=} h^{{A_3}}_{W_1}(\mathbf{r},  W_1\rightarrow W_3), w_2\texttt{=} h^{{A_3}}_{W_2}(\mathbf{r}, W_2 \rightarrow W_3), r_{W_3}).
$$ 

For the first argument of that function call we have $w_1=h^{{A_3}}_{W_1}(\mathbf{r},  W_1\rightarrow W_3) = a_{1,W_1\to W_3} = 1$. Then, for the second argument, 
$
w_2=h^{{A_3}}_{W_2}(\mathbf{r}, W_2 \rightarrow W_3) = a_{2,W_2\to W_3} = 0,
$ 
giving the result $w_3=f_{W_3}(w_1\texttt{=}1, w_2\texttt{=}0, r_{W_3}).$

The procedures for evaluating the functions $g$ and $h^{A_i}$ are sufficient to translate any combined factual and/or potential outcome joint probability statement into probability statements involving only the response function variables $\mathbf{R}$. Thus, using our response function formulation, any potential outcome or factual joint probability statement can be written
%\begin{eqnarray}
%Q_v = \P\{f_{W_{i_{1}}}(f_{W_{i_{1\cdot}}}^{{A}_{i_1}}(\mathbf{R}), R_{W_{i_1}}) = %w_{i_1}, \ldots, f_{W_{i_P}}(f_{W_{i_{P\cdot}}}^{{A}_{i_P}}(\mathbf{R}), %R_{W_{i_P}}) = w_{i_P}, \nonumber \\
%f_{W_{j_1}}(\mathbf{R}) = w_{j_1}, \ldots, f_{W_{j_O}}(\mathbf{R}) = w_{j_O}\}, %\label{cfac}
%\end{eqnarray}
\begin{eqnarray}
Q:=\P\{h_{W_{i_1}}^{A_{i_1}}(\mathbf{R},W_{i_1})=w_{i_1},\dots,h_{W_{i_P}}^{A_{i_P}}(\mathbf{R},W_{i_P})=w_{i_P},\nonumber\\
g_{W_{j_1}}(\mathbf{R})=w_{j_1},\dots,g_{W_{j_O}}(\mathbf{R})=w_{j_O}\}, \label{cfac}
\end{eqnarray}
where $\mathcal{P} = \{i_1, \ldots, i_P\}$ denote the indices of potential outcomes, and $\mathcal{O} = \{j_{1}, \ldots, j_{O}\}$ the indices of the factual outcomes (and these sets may be overlapping).  Given the functional expressions we have defined and our procedures for evaluating them, we can therefore write 
\[
Q=\sum_{\mathbf{r}\in\Gamma(Q)}\P\{\mathbf{R}=\mathbf{r}\}, \mbox{ where}\]
\[\Gamma(Q):=\{\mathbf{r}\in\nu(\mathbf{R}):\forall i_p\in\mathcal{P},w_{i_p}=h^{A_{i_p}}_{W_{i_p}}(\mathbf{r},W_{i_p})\mbox{ and }\forall j_o\in\mathcal{O},w_{j_o}=g_{W_{j_o}}(\mathbf{r})\}.
\]
We will call an expression of this form an \emph{atomic} query. Their form is completely general, and allows arbitrarily nested potential outcomes, and combinations with observational quantities. We will combine atomic queries to obtain causal contrasts of interest, such as the causal risk difference. %Causal contrasts $Q$, such as the risk difference, are constructed by defining $Q$ to be linear combinations of a set of $Q_v$ indexed by $v \in \{1, \ldots, V\}$.

%In other words, for a given probability statement involving factual and/or counterfactual statements, the set $\mathbf{R}$ can be determined by iterating through all possible combinations of $\mathbf{r}$, following the procedures for evaluating the response functionals $f$ for factual outcomes and $f^\mathbf{A}$ for counterfactual outcomes, and identifying the combinations of $\mathbf{r}$ that yields variable values that match with the values $w_{1}, \ldots, w_{n}$.

%\subsection{Obtaining causal query as linear function of causal parameters}

%Let $\hat{r}_1, \ldots, \hat{r}_K$ denote the list of all possible values of the response function variables for the set of variables on the right side, $q_1, \ldots, q_K$ their corresponding probabilities, and the vector of probabilties be denoted $\mathbf{q}$. 

\begin{proposition} \label{t2}
Let $G$ be a causal DAG satisfying Conditions \ref{item:cond1} and \ref{item:cond2}, and let $Q$ be an atomic query satisfying the following Conditions:
\begin{enumerate}
\setcounter{enumi}{4}
\item \label{item:cond3} Each atomic query is a probability as given in Equation \eqref{cfac} where \\
${i_1}, \ldots, {i_P}, {j_1}, \ldots, {j_O} \in \mathcal{R}$ (i.e., all outcome variables must be in $\mathcal{W}_\mathcal{R}$) and 
\item \label{item:cond4} if $\mathcal{L}\ne\varnothing$ then: (i) none of the variables in $\mathcal{W}_\mathcal{L}$ that are intervened upon can have any children in $\mathcal{W}_\mathcal{L}$, (ii) all variables in $\mathcal{W}_\mathcal{L}$ must be in the intervention set, or ancestors of the variables in the intervention set (here the intervention set refers to variables in the rows of the $A$ matrices that are not $\emptyset$), (iii) no observations are allowed, i.e, $\mathcal{O}=\varnothing$. 
\end{enumerate}
Then there exists a constant binary vector $\alpha\in\{0,1\}^{\aleph_\mathcal{R}}$ such that $Q=\alpha^\top\mathbf{q}$.
\end{proposition}

\noindent
See \hyperref[proofoft2]{Appendix A} for proof. A procedure for construction of this $\alpha$ is detailed in Algorithm \ref{queryalgo} which converts the atomic query $Q$ into a binary linear combination of probabilities of response function variables of the $\mathcal{R}$-side. 
\clearpage

\spacingset{1.25}
\begin{algorithm}[H]
    \SetAlgoLined
    \KwResult{$Q$ expressed as a simple sum of a subset of the components of $\mathbf{q}$.}
    Initialize $\alpha\in\{0,1\}^{\aleph_\mathcal{R}}$ by $\forall\gamma\in\{1,\dots,\aleph_\mathcal{R}\}$, $\alpha_\gamma:=1$\;
    Let $\mathcal{P},\mathcal{O}$ be the index sets as defined above corresponding to $Q$\;
    \For{$\gamma\in1,\ldots,\aleph_\mathcal{R}$}{
        \For{$l\in\mathcal{P}$}{
            Construct $A_l$ according to $l$\;
            Compute $\omega := h^{A_l}_{W_l}(\mathbf{r}_\gamma,W_l)$\;
            \If{$\omega\ne w_l$}{
                Set $\alpha_\gamma:=0$\;
                $\mathbf{break}$\;
            }
        }
        \If{$\alpha_\gamma=0$}{
            $\mathbf{break}$\;
        }
        \For{$l \in \mathcal{O}$}{
            Compute $\omega := g_{W_l}(\mathbf{r}_\gamma)$\;
            \If{$\omega\ne w_l$}{
                Set $\alpha_\gamma:=0$\;
                $\mathbf{break}$\;
            }
        }
    }
    %Now $Q=\alpha^T\mathbf{q}$. 
    \caption{Converting $Q$ to a binary linear combination of $\mathbf{q}$. \label{queryalgo}}
\end{algorithm}
\spacingset{2}

\vspace*{1cm}
The following corollary, which specifies the general form of a causal query, follows immediately since linear combinations of linear combinations again are just linear combinations.

\begin{corollary}
Let $Q^*$ be any real linear combination of atomic queries (in particular, $Q$ may be a classic linear causal contrast such as a causal risk difference). Under conditions \ref{item:cond1}, \ref{item:cond2}, \ref{item:cond3}, and \ref{item:cond4},  there exists a constant vector $\alpha^*\in\mathbb{R}^{\aleph_\mathcal{R}}$ such that $Q^*=\alpha^{*\top}\mathbf{q}$.
\end{corollary}

%\subsection{A note on the equivalence class of causal problems for which the bounds are tight}

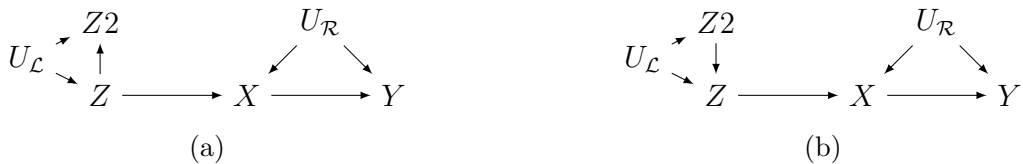
\begin{figure}[h]
\captionsetup[sub]{width=.9\linewidth}
\centering
\resizebox{\linewidth}{!}{
\begin{subfigure}[t]{0.5\textwidth}
\centering
\begin{tikzpicture}
\node (ul) at (-3, .5) {$U_\mathcal{L}$};
\node (z2) at (-2, 1) {$Z2$};
\node (E) at (-2,0) {$Z$};
\node (uh) at (1,1) {$U_\mathcal{R}$};
%\node (u) at (-1,-1) {$V$};
\node (v) at (0,0) {$X$};
\node (i) at (2,0) {$Y$};
\draw[-latex] (E) -- (v);
%\draw [dashed,<->] (u) -- (E);
%\draw [-latex] (u) -- (v);
\draw[-latex] (v) -- (i);
\draw[-latex] (uh) -- (i);
\draw[-latex] (uh) -- (v);
\draw[-latex] (E) -- (z2);
\draw[-latex] (ul) -- (E);
\draw[-latex] (ul) -- (z2);

%\draw[-latex] (uh) -- +(2.79,-0.9) (i);

\end{tikzpicture}
\caption{\label{a1}}
\end{subfigure}

\begin{subfigure}[t]{0.5\textwidth}
\centering
\begin{tikzpicture}
\node (ul) at (-3, .5) {$U_\mathcal{L}$};
\node (z2) at (-2, 1) {$Z2$};
\node (E) at (-2,0) {$Z$};
\node (uh) at (1,1) {$U_\mathcal{R}$};
%\node (u) at (-1,-1) {$V$};
\node (v) at (0,0) {$X$};
\node (i) at (2,0) {$Y$};
\draw[-latex] (E) -- (v);
%\draw [dashed,<->] (u) -- (E);
%\draw [-latex] (u) -- (v);
\draw[-latex] (v) -- (i);
\draw[-latex] (uh) -- (i);
\draw[-latex] (uh) -- (v);
\draw[-latex] (z2) -- (E);
\draw[-latex] (ul) -- (E);
\draw[-latex] (ul) -- (z2);

\end{tikzpicture}
\caption{\label{b1}}
\end{subfigure}
}
\caption{An equivalence class of DAGs defined by arbitrary connections in $\mathcal{W}_\mathcal{L}$. Bounds for causal queries that involve intervening on $X$ that meet our conditions are equivalent and tight for these two graphs in (a) and (b). \label{equiv}}
\end{figure}

The algorithms are formulated so that bounds are derived in terms of the true probabilities of the observed variables in $\mathcal{W}_\mathcal{R}$ conditional on the variables in $\mathcal{W}_\mathcal{L}$. Provided one is not intervening on any of the variables in $\mathcal{W}_\mathcal{L}$, Conditions \ref{item:cond1} and \ref{item:cond2} imply that the directions of the edges within $\mathcal{W}_\mathcal{L}$ cannot influence the bounds. That is, the bounds are tight for the equivalence class of DAGs that contains the set of DAGs for all possible directions of edges among variables in $\mathcal{W}_\mathcal{L}$. For example, the bounds computed for a query such as $\P\{Y(X = 1) = 1\}$ are tight and equal for both of the DAGs in Figures \ref{equiv} (a) and (b). In either case, the knowledge of whether $Z$ causes $Z2$ or vice versa does not influence the bounds because both of those variables are conditioned upon in the algorithm. 

Alternatively, if the desired query was $\P\{Y(X(Z = 1)) = 1\}$, the DAGs in Figures \ref{equiv} (a) and (b) may not result in the same bounds, and in fact, the causal problem under Figure \ref{equiv} (a) may not be linear. As required by Conditions \ref{item:cond3} and \ref{item:cond4}, if we intervene upon a variable in $\mathcal{W}_\mathcal{L}$, then the direction of edges within $\mathcal{W}_\mathcal{L}$ matters, and in fact if the intervened upon variable has a child also in $\mathcal{W}_\mathcal{L}$, the condition will not be met.

\section{Optimization via vertex enumeration} \label{optimizationSection}

After applying Algorithms \ref{probalgo} and \ref{queryalgo}, we have a linear objective and a system of linear constraints. We also have the probabilistic constraints: 
\begin{eqnarray*}
\forall\mathbf{w}_\mathcal{L}\in\nu(\mathbf{W}_\mathcal{L}),\quad \sum_{\mathbf{w}_\mathcal{R}\in\nu(\mathbf{W}_\mathcal{R})}\P\{\mathbf{W}_\mathcal{R}=\mathbf{w}_\mathcal{R}\mid\mathbf{W}_\mathcal{L}=\mathbf{w}_\mathcal{L}\}=1\\
\end{eqnarray*}
 and 
\begin{eqnarray*}
\sum_{\gamma=1}^{\aleph_\mathcal{R}}q_\gamma=\sum_{\gamma=1}^{\aleph_\mathcal{R}}\P\{\mathbf{R}_\mathcal{R}=\mathbf{r}_\gamma\}=\sum_{\mathbf{r}_\mathcal{R}\in\nu(\mathbf{R}_\mathcal{R})}\P\{\mathbf{R}_\mathcal{R}=\mathbf{r}_\mathcal{R}\}=1.
\end{eqnarray*}
Additional linear constraints on ${\mathbf{q}}$ can be optionally given as $B\mathbf{q}\ge\mathbf{d}$ where $B$ and $\mathbf{d}$ are respectively a matrix and vector of real constants. These constraints can be used to encode assumptions about the response functions that are not possible to encode in a DAG, for example, restricting the probabilities of implausible response patterns. We thus arrive at the following linear programming problem for the lower bound; the upper bound is given by the corresponding maximization problem. 

\begin{eqnarray*}
\text{minimize }Q&=\alpha^T\mathbf{q}\\
\text{subject to }P\mathbf{q}&=\mathbf{p},\\
B\mathbf{q}&\ge\mathbf{d},\\
\mathbf{q}&\geq\mathbf{0},\\
\text{and }\mathbf{1}^T\mathbf{q}&=1
\end{eqnarray*}

Note that the constraint space constitutes a bounded (due to the probabilistic constraints) convex polytope. By the fundamental theorem of linear programming, the global extrema must occur at one of the vertices of the polytope. We can thus solve this problem symbolically by applying an efficient vertex enumeration algorithm, such as the double description algorithm \citep{doubledescriptionmethod, cdd} to enumerate the vertices of the polytope of the dual linear program. For instance, the dual of the minimization problem above is given by 

\begin{eqnarray*}
\text{maximize }
&\begin{pmatrix}\mathbf{d}^T&1&\mathbf{p}^T\end{pmatrix}\mathbf{y}\\
\text{subject to }
&\begin{pmatrix}
    B^T & \begin{matrix}1 & P^T\end{matrix}\\
    I & 0
\end{pmatrix}
\mathbf{y}\le
\begin{pmatrix}
    \alpha\\\mathbf{0}
\end{pmatrix}.
\end{eqnarray*}
So by the strong duality theorem, the optimum of the dual, and thus also of the primal problem, is of the form  $\begin{pmatrix}\mathbf{d}^T&1&\mathbf{p}^T\end{pmatrix}\bar{\mathbf{y}}$ where $\bar{\mathbf{y}}$ is a vertex of the polytope $\{\mathbf{y}:\begin{pmatrix}
    B^T & \begin{matrix}1 & P^T\end{matrix}\\
    I & 0
\end{pmatrix}
\mathbf{y}\le
\begin{pmatrix}
    \alpha\\\mathbf{0}
\end{pmatrix}\}$.
This gives a lower bound on the causal effect of interest as the maximum of a set of expressions involving only observable probabilities. Similarly, the upper bound is given by reversing the dual inequality and minimizing over the corresponding polytope. 

\begin{proposition} \label{tightness}
Under conditions \ref{item:cond1}-\ref{item:cond4} and subject to any additional linear constraints of the form $B\mathbf{q}\ge\mathbf{d}$, the procedure above yields valid and {tight} symbolic bounds for a causal query that is a linear combination of atomic queries. 
\end{proposition}

\begin{corollary}
If condition \ref{item:cond6} does not hold, then the bounds derived using the above procedure are still valid. 
\end{corollary}

See \hyperref[proofoftightness]{Appendix A} for proof. The conditions \ref{item:cond5} and \ref{item:cond6} represent a worst-case scenario of confounding and ensure that the decompositions giving rise to the linear constraints cannot be further factorized to yield more granular but non-linear constraints. If however there is any known (partial) absence of such confounding, then these bounds are still valid, and may be narrow enough to be informative, while not necessarily tight. Such an absence of confounding on the $\mathcal{R}$-side implies some independence among the $\mathbf{R}_\mathcal{R}$ variables, and hence additional constraints on their distribution. Thus the true feasible space may be smaller than the one considered in our algorithm, but completely contained inside it.

%If assumptions (\ref{item:cond5}) and (\ref{item:cond6}) hold, then the decompositions giving rise to the linear constraints cannot be further factorized to yield more granular but non-linear constraints. Hence, for this class of problems, the procedure above yields valid and \emph{tight} symbolic bounds. If however there is any \emph{known} (partial) \emph{absence} of such confounding, then these bounds are still \emph{valid}, and may be narrow enough to be informative, while not necessarily tight. 

\section{Examples} \label{examplesSection}
%\subsection{User interface}

%The \texttt{R} package \texttt{causaloptim}, available now on CRAN, has a graphical user interface which allows users to define a DAG that is constrained by design to be in the class of problems that is characterized above \citep{arr,causaloptim}. 
The graphs in the following examples are divided into a left side, which corresponds to the $\mathcal{W}_\mathcal{L}$ set, and a right side, which corresponds to the $\mathcal{W}_\mathcal{R}$ set, as in Figure \ref{base1}. The left side is displayed as a violet (dark grey) box, and the right side a yellow (light grey) box. %The program is constrained so that only DAGs that meet Assumptions 1-4 may be drawn. Unmeasured common causes on each side are added to the DAG automatically by the program.

%After a DAG interactively using a web browser, they specify the causal effect of interest using the same notation we have used in this paper in a text interface. Additional user-specified constraints are optional and are also specified using a text interface. The program then applies our algorithm in order to find the symbolic bounds in terms of the observed conditional probabilities.

%In cases where there is only a single intervention set that applies for all possible paths, the program allows for a shorthand notation. For example, in the graph in Figure \ref{b}, if the query of interest is $\P\{W_3(W_2(W_1 = 1), W_1 = 1) = 1\}$, then the user may instead write $\P\{W_3(W_1 = 1) = 1\}$. This is understood by the program to indicate a single intervention set on $W_1$, and the interventions are propagated through all possible paths to the outcome $W_3$. This is useful in situations where there may be a large number of possible paths between a single intervention of interest and the outcome.  

\subsection{Confounded exposure and outcome}

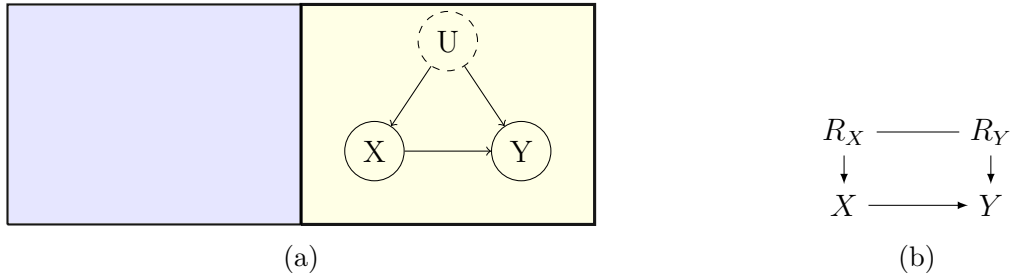
\begin{figure}[ht]
\captionsetup[sub]{width=.9\linewidth}
\centering
\resizebox{\linewidth}{!}{
\begin{subfigure}[t]{0.5\textwidth}
\centering
\begin{tikzpicture}
	 \tikzstyle{NodeStyle} = [shape = circle, minimum width = 6ex]
		
	\draw[fill=rightcol, opacity=0.1, very thick] (3,-1) -- (3,2) -- (7,2) -- (7,-1) -- (3,-1);
	\draw[fill=leftcol, opacity=0.1, very thick] (-1,-1) -- (3,-1) -- (3,2) -- (-1,2) -- (-1,-1);

		\draw[opacity=0.9, thick] (-1,-1) -- (3,-1) -- (3,2) -- (-1,2) -- (-1,-1);
		
	\draw[opacity=0.9, very thick] (3,-1) -- (3,2) -- (7,2) -- (7,-1) -- (3,-1);
		
		\tikzstyle{EdgeStyle}   = [->,>=stealth']

        \node[draw, circle] at (4, 0)   (2) {X};
        \node[draw, circle] at (6, 0)   (3) {Y};
         \node[draw, circle, dashed] at (5, 1.5)  (4) {U};

	\draw [->] (2) to (3);
    \draw [->] (4) to (3);
    \draw [->] (4) to (2);
\end{tikzpicture}
\caption{\label{base1}}
\end{subfigure}

\begin{subfigure}[t]{0.5\textwidth}
\centering
\begin{tikzpicture}

\node (rx) at (0,1) {$R_X$};
\node (ry) at (2,1) {$R_Y$};
\node (x) at (0,0) {$X$};
\node (y) at (2,0) {$Y$};
\draw[-latex] (x) -- (y);
\draw[-latex] (rx) -- (x);
\draw[-latex] (ry) -- (y);
\draw[-latex, -] (rx) -- (ry);

\end{tikzpicture}
\caption{\label{base2}}
\end{subfigure}
}
\caption{Simple confounded example and the equivalent response function variable graph. }
\end{figure}

The basic DAG with two variables that are confounded as shown in Figure \ref{base1} conforms to our class of models. In this case, the variable $X$ is the exposure of interest, and $Y$ the outcome of interest. $X$ and $Y$ have a common, unmeasured cause $U$. We specify $X$ and $Y$ to be ternary and binary respectively, so $X$ takes values in $\{0,1,2\}$ and $Y$ in $\{0,1\}$. Our causal effects of interest are the risk differences $\P\{Y(X = 2) = 1\} - P\{Y(X = 0) = 1\}, \P\{Y(X = 2) = 1\} - P\{Y(X = 1) = 1\}\text{ and }\P\{Y(X = 1) = 1\} - P\{Y(X = 0) = 1\}$, and we have no additional constraints to specify. 

Here we have two variables and therefore two response function variables. The response function variable formulation of the graph in Figure \ref{base2} is an equivalent representation of the causal model. The following tables define the values of the response functions and variables: 

\vspace{0.5cm}
\begin{tabular}{c|c}
    $x=f_X(r_X)$\\
    \hline
    $r_X=0$&$x=0$\\
    $r_X=1$&$x=1$\\
    $r_X=2$&$x=2$\\
\end{tabular}
\begin{tabular}{c|ccc}
    $y=f_Y(x,r_Y)$&$x=0$&$x=1$&$x=2$\\
    \hline
    $r_Y=0$&$y=0$&$y=0$&$y=0$\\
    $r_Y=1$&$y=1$&$y=0$&$y=0$\\
    $r_Y=2$&$y=0$&$y=1$&$y=0$\\
    $r_Y=3$&$y=1$&$y=1$&$y=0$\\
    $r_Y=4$&$y=0$&$y=0$&$y=1$\\
    $r_Y=5$&$y=1$&$y=0$&$y=1$\\
    $r_Y=6$&$y=0$&$y=1$&$y=1$\\
    $r_Y=7$&$y=1$&$y=1$&$y=1$\\
\end{tabular}
\vspace{0.5cm}

$R_X$ is a random variable that can take on $3$ possible values, and $R_Y$ is a random variable that can take on $2^3=8$ possible values. Thus, the joint distribution of $(R_X, R_Y)$ is characterized by $3\cdot8=24$ parameters, say $q_{i, j}$, where $i \in \{0, 1, 2\}$ and $j \in \{0, 1, 2, 3, 4, 5, 6, 7\}$. 
Applying Algorithm \ref{probalgo}, we can relate the $3\cdot2=6$ observed probabilities to the parameters of the response function variable distribution as follows: 

\begin{eqnarray*}
p_{0,0;}:=\P\{X = 0,Y = 0\} &=& q_{0,0} + q_{0,2} + q_{0,4} + q_{0,6}\\
p_{1,0;}:=\P\{X = 1,Y = 0\} &=& q_{1,0} + q_{1,1} + q_{1,4} + q_{1,5}\\
p_{2,0;}:=\P\{X = 2,Y = 0\} &=& q_{2,0} + q_{2,1} + q_{2,2} + q_{2,3}\\
p_{0,1;}:=\P\{X = 0,Y = 1\} &=& q_{0,1} + q_{0,3} + q_{0,5} + q_{0,7}\\
p_{1,1;}:=\P\{X = 1,Y = 1\} &=& q_{1,2} + q_{1,3} + q_{1,6} + q_{1,7}\\
p_{2,1;}:=\P\{X = 2,Y = 1\} &=& q_{2,4} + q_{2,5} + q_{2,6} + q_{2,7}.
\end{eqnarray*}

We get 
\begin{eqnarray*}
A & = & \left[\begin{array}{c|cc}
& X\rightarrow Y & Y\\
\hline
X&0&\emptyset\\
Y&\emptyset&\emptyset  \\
\end{array} \right], \mbox{ for }\P\{Y(X = 0) = 1\}, \\ 
A & = & \left[\begin{array}{c|cc}
& X\rightarrow Y & Y\\
\hline
X&1&\emptyset\\
Y&\emptyset&\emptyset  \\
\end{array} \right], \mbox{ for }\P\{Y(X = 1) = 1\}\mbox{ and} \\
A & = & \left[\begin{array}{c|cc}
& X\rightarrow Y & Y\\
\hline
X&2&\emptyset\\
Y&\emptyset&\emptyset  \\
\end{array} \right], \mbox{ for }\P\{Y(X = 2) = 1\}.
\end{eqnarray*}

Applying Algorithm \ref{queryalgo}, we get 

\begin{align*}
    \P\{Y(X=0)=1\}
    &=q_{0,1}+q_{0,3}+q_{0,5}+q_{0,7}\\
    &+q_{1,1}+q_{1,3}+q_{1,5}+q_{1,7}\\
    &+q_{2,1}+q_{2,3}+q_{2,5}+q_{2,7},\\
    \P\{Y(X=1)=1\}
    &=q_{0,2}+q_{0,3}+q_{0,6}+q_{0,7}\\
    &+q_{1,2}+q_{1,3}+q_{1,6}+q_{1,7}\\
    &+q_{2,2}+q_{2,3}+q_{2,6}+q_{2,7}\text{ and }\\
    \P\{Y(X=2)=1\}
    &=q_{0,4}+q_{0,5}+q_{0,6}+q_{0,7}\\
    &+q_{1,4}+q_{1,5}+q_{1,6}+q_{1,7}\\
    &+q_{2,4}+q_{2,5}+q_{2,6}+q_{2,7},
\end{align*}

hence the contrasts 
\begin{align*}
\P\{Y(X=1)=1\}-\P\{Y(X=0)=1\}&=q_{0,2}+q_{0,6}+q_{1,2}+q_{1,6}+q_{2,2}+q_{2,6}\\&-q_{0,1}-q_{0,5}-q_{1,1}-q_{1,5}-q_{2,1}-q_{2,5},\\
\P\{Y(X=2)=1\}-\P\{Y(X=0)=1\}&=q_{0,4}+q_{0,6}+q_{1,4}+q_{1,6}+q_{2,4}+q_{2,6}\\&-q_{0,1}-q_{0,3}-q_{1,1}-q_{1,3}-q_{2,1}-q_{2,3}\text{ and }\\
\P\{Y(X=2)=1\}-\P\{Y(X=1)=1\}&=q_{0,4}+q_{0,5}+q_{1,4}+q_{1,5}+q_{2,4}+q_{2,5}\\&-q_{0,2}-q_{0,3}-q_{1,2}-q_{1,3}-q_{2,2}-q_{2,3}.
\end{align*}

Together with the probabilistic constraints, we then have the fully specified linear programming problem. The bounds as output by the program are 

\[\begin{array}{cc}
    &\P\{X=0,Y=0\}+\P\{X=1,Y=1\}-1\\
    &\le\P\{Y(X=1)=1\}-\P\{Y(X=0)=1\}\le\\
    &1-\P\{X=1,Y=0\}-\P\{X=0,Y=1\},
\end{array}\]
\[\begin{array}{cc}
    &-\P\{X=1,Y=0\}-\P\{X=2,Y=0\}-\P\{X=0,Y=1\}-\P\{X=1,Y=1\}\\
    &\le\P\{Y(X=2)=1\}-\P\{Y(X=0)=1\}\le\\
    &1-\P\{X=2,Y=0\}-\P\{X=0,Y=1\}
\end{array}\]
and
\[\begin{array}{cc}
    &-\P\{X=0,Y=0\}-\P\{X=2,Y=0\}-\P\{X=0,Y=1\}-\P\{X=1,Y=1\}\\
    &\le\P\{Y(X=2)=1\}-\P\{Y(X=1)=1\}\le\\
    &1-\P\{X=2,Y=0\}-\P\{X=1,Y=1\}.
\end{array}\]

\subsection{Two instruments}

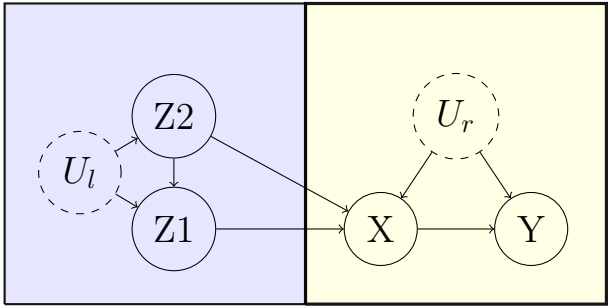
\begin{figure}[h]
	\large{\begin{tikzpicture}
	 \tikzstyle{NodeStyle} = [shape = circle, minimum width = 6ex] % draw]
		
	\draw[fill=rightcol, opacity=0.1, very thick] (3,-1) -- (3,3) -- (7,3) -- (7,-1) -- (3,-1);
	\draw[fill=leftcol, opacity=0.1, very thick] (-1,-1) -- (3,-1) -- (3,3) -- (-1,3) -- (-1,-1);
	\draw[opacity=0.9, thick] (-1,-1) -- (3,-1) -- (3,3) -- (-1,3) -- (-1,-1);
	\draw[opacity=0.9, very thick] (3,-1) -- (3,3) -- (7,3) -- (7,-1) -- (3,-1);

		\tikzstyle{EdgeStyle}   = [->,>=stealth']

		\node[draw, circle] at (1.25, 0)   (1) {Z1};
		\node[draw, circle] at (1.25, 1.5)   (2) {Z2};
        \node[draw, circle] at (4, 0)   (3) {X};
        \node[draw, circle] at (6, 0)   (4) {Y};
         \node[draw, circle, dashed] at (5, 1.5)  (5) {$U_r$};
         \node[draw, circle, dashed] at (0, .75)  (6) {$U_l$};

	\draw [->] (2) to (1);
	\draw [->] (2) to (3);
	\draw [->] (1) to (3);
    \draw [->] (3) to (4);
    \draw [->] (5) to (3);
    \draw [->] (5) to (4);
    \draw [->] (6) to (1);
    \draw [->] (6) to (2);

	\end{tikzpicture}}
	\caption{Two instrumental variables example with binary variables\label{twoiv}}
\end{figure}

Our next example is shown in the DAG in Figure \ref{twoiv}. This extends the instrumental variable example to the case where there are two binary variables on the left side that may be associated with each other and that both have a direct effect on $X$, but no direct effect on $Y$. This situation may arise in Mendelian randomization studies, wherein multiple genes may be known to cause changes in an exposure but not directly on the outcome.

The bounds on risk difference $\P\{Y(X = 1)\} - \P\{Y(X = 0)\}$ under this DAG can be computed using our method. In this problem, there are 16 constraints involving the conditional probabilities, the distribution of the response function variables of the $\mathcal{R}$-side has 64 parameters, and the causal query is a function of 32 of these parameters. The bounds are the extrema over 112 vertices, and are therefore too long to be presented simply, but they are included in the Supplementary Material along with code to reproduce the results using our method. %This causal problem is equivalent to having a single 4-level instrumental variable.

To illustrate these bounds, we computed them for specific values of observed probabilities generated from the model in Equation \eqref{eq:simmod} which satisfies the DAG in Figure \ref{twoiv}. Using these simulations we compare our bounds to the classic IV bounds from \citet{balke1997bounds} for a single binary instrument and to bounds derived using our method for a single but 4-level categorical instrument. 

For each of 50,000 simulations, we generated values $pu_l$ and $pu_l$ of probabilities of the latent influences $U_l$ and $U_r$ from the standard uniform distribution, and each of 12 parameters $\alpha_1,\alpha_2,\alpha_3,\alpha_4,\alpha_5,\beta_1,\beta_2,\beta_3,\beta_4,\gamma_1,\gamma_2,\gamma_3$ from the normal distribution with mean 0 and standard deviation 2. Assuming that the conditional distributions of the observed variables follow probit models, we can derive, by Bayesian decomposition according to the diagram in Figure \ref{twoiv}, the joint distribution of $\P(U_l,U_r,Z1,Z2,X,Y)$. From that, we marginalize out the variables $U_l$ and $U_r$ to get $\P\{Z1,Z2,X,Y\}$ and finally compute and divide this by the marginal joint probability $\P\{Z1,Z2\}$ of the instruments $Z1$ and $Z2$, to get the conditional probability distribution $\P\{X,Y|Z1,Z2\}$ that goes into the symbolic expressions of the tight bounds. We do a similar marginalization of $Z2$ in order to get conditional probabilities $\P\{X,Y|Z1\}$ for computation of the single binary IV bounds. In each simulation, we create values of probabilities $\P\{Z3=z3\},z3\in\{0,1,2,3\}$ of a 4-level instrument $Z3$ from probabilities $\P\{Z1=z1,Z2=z2\},z1,z2\in\{0,1\}$ to get appropriate input for the expressions of the tight bound computed in the single 4-level instrument setting.

\begin{figure}[ht]
\begin{center}
    \includegraphics[width= .99\textwidth]{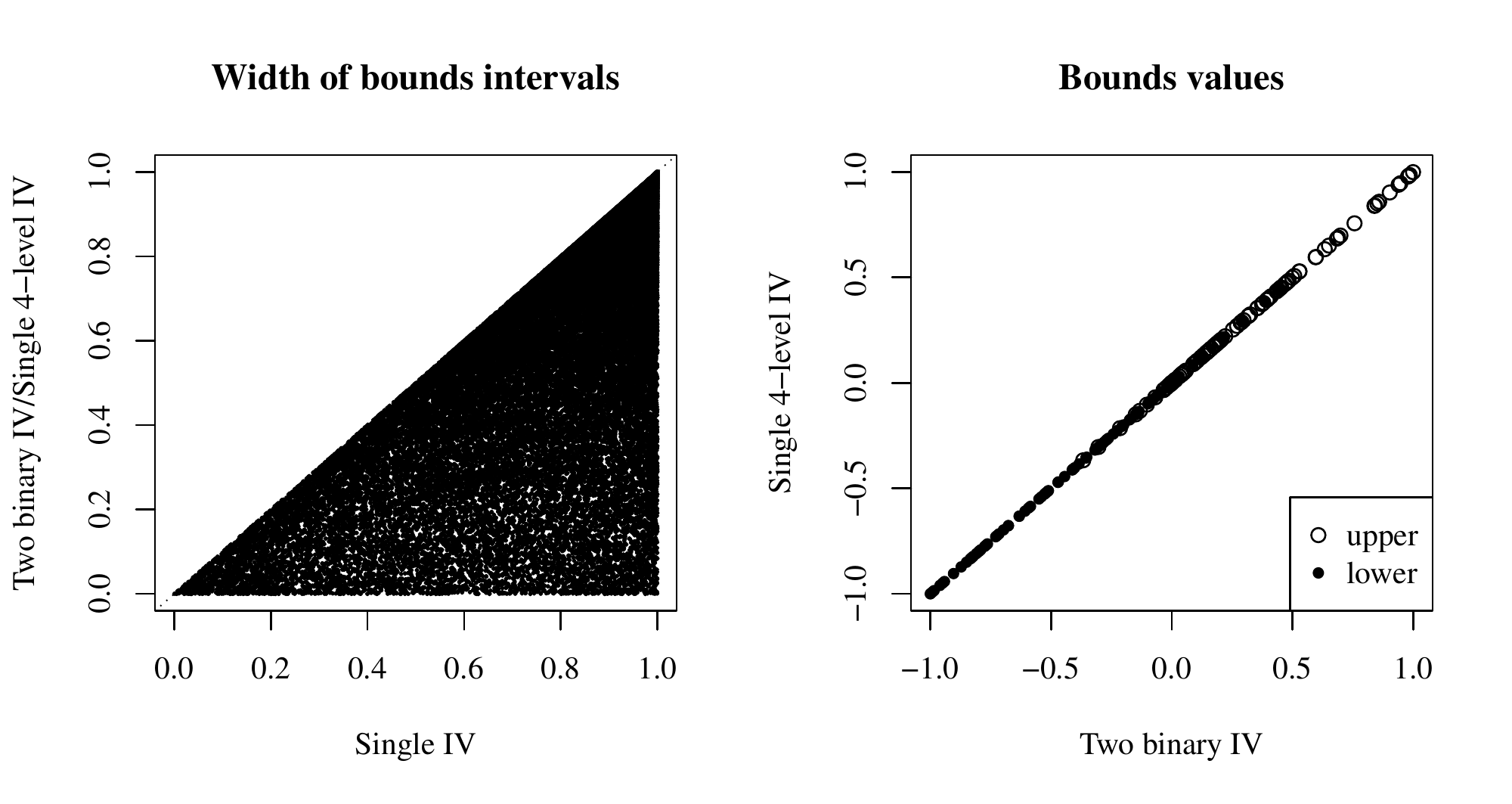}
\end{center}
\caption{Under a DAG with two instruments, the left panel is a comparison of the width of the bounds intervals for the causal risk difference assuming only one of the instruments is observed to the width of the bounds assuming both are observed. The right panel compares the values of the upper and lower bounds for each replicate for two binary instruments versus a single 4-level instrument. \label{compb}}
\end{figure}

The the widths of the classic IV bounds and the dual binary instruments are compared in Figure \ref{compb}. The bounds with two
instruments are never wider than the classic IV bounds with a single binary instrument. The simulations also verify that a single four level instrument yields exactly the same bounds as two binary ones. \texttt{R} code for these simulations are provided in the Supplementary Material.

%Specifically, we generated probability distributions $\P\{U_l, U_r,Z1, Z2,X,Y\}$ under the causal diagram in Figure \ref{twoiv} from the model

\begin{eqnarray}
\label{eq:simmod}
\begin{array}{rcl}
\P\{U_l=1\} & \sim & \mbox{Unif}(0,1) \\
\P\{U_r=1\} & \sim & \mbox{Unif}(0,1) \\
\P\{Z2=1|U_l\} & = & \Phi(\alpha_1+\alpha_2U_l) \\
\P\{Z1=1|U_l, Z2\} & = & \Phi(\alpha_3+\alpha_4U_l + \alpha_5Z2) \\
\P\{X=1|U_r, Z1, Z2\} & = & \Phi(\beta_1+\beta_2U_r + \beta_3Z1 + \beta_4Z2) \\
\P\{Y=1|U_r, X\} & = & \Phi(\gamma_1+\gamma_2U_r + \gamma_3X) \\
(\alpha_1,\alpha_2,\alpha_3,\alpha_4,\alpha_5,\beta_1,\beta_2,\beta_3,\beta_4,\gamma_1,\gamma_2, \gamma_3) & \sim & N(0,4)
\end{array}
\end{eqnarray}
%where $\Phi(x)$ is the cumulative distribution function of a standard normal random variable. 

%The results are shown in Figure \ref{compb} for 5,000 simulated distributions. The bounds with two instruments are never wider than those with only one instrument. 

\clearpage

\subsection{Measurement error in the outcome}

\begin{figure}[h]
	\large{\begin{tikzpicture}
	 \tikzstyle{NodeStyle} = [shape = circle, minimum width = 6ex] % draw]
		
	\draw[fill=rightcol, opacity=0.1, very thick] (3,-1) -- (3,3) -- (7,3) -- (7,-1) -- (3,-1);
	\draw[fill=leftcol, opacity=0.1, very thick] (-1,-1) -- (3,-1) -- (3,3) -- (-1,3) -- (-1,-1);
	\draw[opacity=0.9, thick] (-1,-1) -- (3,-1) -- (3,3) -- (-1,3) -- (-1,-1);
	\draw[opacity=0.9, very thick] (3,-1) -- (3,3) -- (7,3) -- (7,-1) -- (3,-1);

		\tikzstyle{EdgeStyle}   = [->,>=stealth']

        \node[draw, circle] at (2, 0)   (1) {X};
        \node[draw, circle] at (6, 0)   (2) {Y2};
        \node[draw, circle, dashed] at (4, 0)   (3) {Y};
         \node[draw, circle, dashed] at (5, 1.5)  (4) {$U_r$};
         \node[draw, circle, dashed] at (0, .75)  (5) {$U_l$};

	\draw [->] (1) to (3);
	\draw [->] (3) to (2);
	\draw [->] (4) to (2);
    \draw [->] (4) to (3);
    \draw [->] (5) to (1);
    
	\end{tikzpicture}}
	\caption{Example with measurement error in the outcome. Dashed circles indicate unobserved variables.\label{measerror}}
\end{figure}
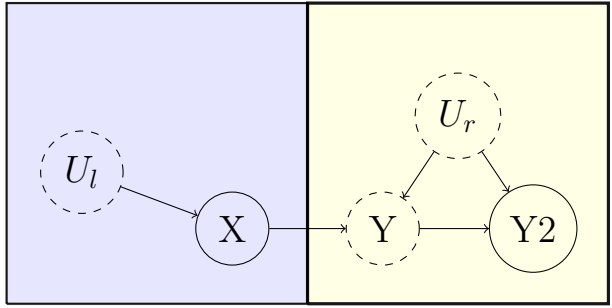

Our final example illustrates some additional features of our method. In Figure \ref{measerror}, we have a binary variable $X$ affecting a binary variable $Y$, but $Y$ is not observed. Instead, the binary variable $Y2$ which is a child of $Y$ is observed, and the effect of the true $Y$ on the measured $Y2$ is confounded. Additionally, we would like to include a constraint that $Y2(Y = 1) \geq Y2(Y = 0)$, which is often called the monotonicity constraint. This constraint encodes the assumption that the outcome measured with error would not be equal to 0 unless the true unobserved outcome is also equal to 0. In terms of the response functions, this constraint removes the case where $f_{Y2}(y, r_{Y2}) = 1 - y$, thereby reducing the number of possible values that $r_{Y2}$ can take by 1. 

The fact that $Y$ is unobserved implies that we have 4 possible conditional probabilities to work with; $\P\{Y2 = y2 | X = x\}$, for $y2, x \in \{0, 1\}$. There are 12 parameters that characterize the distribution of the response function variables of the $\mathcal{R}$-side, and 4 constraints involving conditional probabilities. The bounds for the risk difference $\P\{Y(X = 1) = 1\} - \P\{Y(X = 0) = 1\}$ derived using our method are given by

\begin{eqnarray*}
\mbox{max}\{- 1, 2\, \P\{Y2 = 0 | X = 0\} - 2\, \P\{Y2 = 0 | X = 1\}  - 1 \}\\
\leq \P\{Y(X = 1) = 1\} - \P\{Y(X = 0) = 1\} \leq \\
\mbox{min}\{1, 2\, \P\{Y2 = 0 | X = 0\} - 2\, \P\{Y2 = 0 | X = 1\} + 1\}.
\end{eqnarray*}
Except in cases where $\P\{Y2 = 0 | X = 0\} = \P\{Y2 = 0 | X = 1\}$, these bounds are informative; meaning they give an interval that is shorter than the a priori interval $[-1, 1]$.

\newpage

\section{Conclusion and Discussion}
We have described a general method for the symbolic computation of bounds on causal queries that are not identified from the true probability distribution of the observed variables. For this method, we give two algorithms for deriving the needed constraints and objective to construct such bounds. We describe a class of causal graphs and queries that will always define a linear program, for which we have shown the derived symbolic bounds will always be both valid and tight. We also show that under a broader class of problems our method will provide valid and possibly informative bounds that are not guaranteed to be tight.

%This has been implemented in the \texttt{R} \citep{arr} package \texttt{causaloptim} with a user-friendly interface that allows for graphical input of causal DAGs and queries and optionally additional constraints in a natural way \citep{causaloptim}. All in a web browser, users can draw DAGs, define causal targets, impose constraints, compute bounds, and output them as text, \LaTeX \ formulas, or \texttt{R} functions. Advanced users can interface with the algorithm directly using code, to ensure reproducibility or for more complex situations. 

Our approach is useful in several novel scenarios, as illustrated in the examples above. Additional applications of this method to unsolved problems in causal inference are now much more accessible to researchers as a class of problems for which linear programming can always be used is well-defined and clear algorithms exist for translating DAGs plus causal queries into linear programs. Our representation of causal estimands as arbitrarily nested counterfactuals and our procedure for translating them into functional expressions provides a significant advance over previous methods. This allows for bounding of cross-world counterfactual quantities which are highly relevant in mediation settings. The generality yet accessibility of the method all but guarantees that practitioners will find novel applications that we have not forseen.

Although our class of problems and method from deriving bounds puts no limit of the number of variables or categories for a given variable, in practice attention must be paid to computational complexity.
%Our basic example and previously described bounds such as the instrumental variable \citep{balke1994counterfactual}, controlled direct effect \citep{cai2008bounds}, natural direct effect \citep{sjolander2014bounds} and multiple instrumental variable problem all run in a matter of seconds using our software on a modern laptop computer. 
%The multiple instrumental variable problem takes approximately 6 hours, which involved enumerating 112 vertices twice (once for the upper bound and once for the lower). 
%There is no theoretical upper limit to the number of vertices that can be enumerated using this approach. %Modern vertex enumeration algorithms and implementations using parallel processing may allow currently unfeasible problems to be solved. 
Since we have $|\nu(R_{W_i})|=\prod_{i=1}^{n}c_{W_i}^{\prod_{V\in\mathbf{pa}_{W_i}}c_V}$ for each variable $W_i$, the cardinalities of the domains of the response function variables grow exponentially with the those of other variables in the DAG. The exact growth pattern will of course depend on the DAG and its connectivity as well as the number of categorical levels of select influential variables. Thus, the number of variables or levels may be limited by computing power.  %Investigation into the computational limits of this approach is ongoing, but even for quite simple settings, keeping the levels below say $5$ would be well-advised.

It should be noted that our conditions for a class of problems to be linear are sufficient, but not necessary. Thus, we cannot rule out that there exist problems outside of our class that can be stated as linear. It may be possible to identify a broader class of problems or a different algorithm that may apply on a case-by-case basis. Nonlinear causal queries such as the relative risk or odds ratio yield nonlinear optimization problems yet in some cases it may be possible to translate them to equivalent linear problems. Measured confounding, or knowledge about the absence of confounding often implies nonlinear constraints. We have assumed that all variables are categorical, although many real scientific problems involve continuous variables. Extensions and insights into solving these sorts of problems would be useful in the causal inference community and are areas of future research for the authors.

\section*{Supplemental material}

Supplementary Material available online includes additional and more detailed results for the two instruments example. The \texttt{R} package \texttt{causaloptim}: An Interface to Specify Causal Graphs and Compute Bounds on Causal Effects, is available from CRAN, and from Github at \if0\blind{[blinded url]}\fi \if1\blind{\url{https://sachsmc.github.io/causaloptim}}\fi, with additional documentation and examples. The file \texttt{example-code.R} contains the R code used to run the examples and simulations presented in the main text.

\bibliographystyle{abbrvnat}
\bibliography{main}

\section*{Appendix A} \label{appendixA}

\begin{proof}[Proof of Proposition~\ref{responsefunctionvariables}]
\label{proofprop1}
For each $W\in\mathcal{W}$, if $\phi_W:\nu(U_W)\to\{h:\nu(\mathbf{Pa}_W)\to \nu(W)\}$ is given by $u_W\mapsto h_{u_W}$, where the \emph{response function} $h_{u_W}$ is given by $\mathbf{pa}_W\mapsto F_W(\mathbf{pa}_W,u_W)$, then let the set of values of the \emph{response function variable} $R_W$ corresponding to $W$,  $\nu(R_W):=\nu(U_W)/\phi_W$ be the partition of $\nu(U_W)$ induced by the equivalence relation $u_1\sim u_2:\iff\phi_W(u_1)=\phi_W(u_2)$. $\phi_W$ maps $\nu(U_W)$ bijectively to the finite set $\{h:\nu(\mathbf{Pa}_W)\to\nu(W)\}$ of response functions. Thus, for each $u_W\in\nu(U_W)$ there exists a unique $r_W\in\nu(R_W)$ and $f_W(\cdot, r_W) \in \{h:\nu(\mathbf{Pa}_W)\to\nu(W)\}$ such that $F_W(\cdot,u_W)=f_W(\cdot,r_W)$. We will henceforth refer to $f_W(\cdot,r_W)$ as the response function and $R_W$ the response function variable. Note that the set $\{h:\nu(\mathbf{Pa}_W)\to\nu(W)\}$ is finite with cardinality $|\nu(W)|^{|\nu(\mathbf{Pa}_w)|}$ since $|\nu(W)|$ and $|\nu(\mathbf{Pa}_w)|$ are both finite.
\end{proof}

\begin{proof}[Proof of Proposition~\ref{linprobth}]
\label{proofoflinprobth}

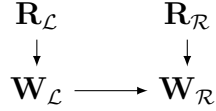
\begin{figure}[h]
\centering
\centering
\begin{tikzpicture}
\node (WL) at (0,0) {$\mathbf{W}_\mathcal{L}$};
\node (RL) at (0,1) {$\mathbf{R}_\mathcal{L}$};
\node (WR) at (2,0) {$\mathbf{W}_\mathcal{R}$};
\node (RR) at (2,1) {$\mathbf{R}_\mathcal{R}$};
\draw[-latex] (RL) -- (WL);
\draw[-latex] (WL) -- (WR);
\draw[-latex] (RR) -- (WR);
\end{tikzpicture}

\caption{A birds-eye view of $G$ in Proposition~\ref{linprobth}. $G$ yields the Bayesian decomposition $\P\{\mathbf{W}_\mathcal{L}=\mathbf{w}_\mathcal{L},\mathbf{W}_\mathcal{R}=\mathbf{w}_\mathcal{R},\mathbf{R}_\mathcal{L}=\mathbf{r}_\mathcal{L},\mathbf{R}_\mathcal{R}=\mathbf{r}_\mathcal{R}\}=\P\{\mathbf{W}_\mathcal{L}=\mathbf{w}_\mathcal{L}\mid \mathbf{R}_\mathcal{L}=\mathbf{r}_\mathcal{L}\}\P\{\mathbf{R}_\mathcal{L}=\mathbf{r}_\mathcal{L}\}\P\{\mathbf{W}_\mathcal{R}=\mathbf{w}_\mathcal{R}\mid \mathbf{W}_\mathcal{L}=\mathbf{w}_\mathcal{L},\mathbf{R}_\mathcal{R}=\mathbf{r}_\mathcal{R}\}\P\{\mathbf{R}_\mathcal{R}=\mathbf{r}_\mathcal{R}\}$.\label{theorem1DAGoverview}}
\end{figure}

Conditions \ref{item:cond1} and \ref{item:cond2} are depicted in Figure \ref{theorem1DAGoverview}. 
Note that this illustrates the setting at a macro-level only, and indicates only the independence relations between the vector-valued variables $\mathbf{W}_\mathcal{L},\mathbf{W}_\mathcal{R},\mathbf{R}_\mathcal{L}\text{ and }\mathbf{R}_\mathcal{R}$ at this level. The internal dependencies among the component variables of $\mathbf{W}_\mathcal{L}$ and $\mathbf{W}_\mathcal{R}$ are further given by the actual "fine-grained" DAG $G$. Regarding the internal dependencies among the component variables of the latent $\mathbf{R}_\mathcal{L}$ and $\mathbf{R}_\mathcal{R}$, we make no assumptions whatsoever, which amounts to assuming potential mutual dependency among all component variables within $\mathbf{R}_\mathcal{L}$ and $\mathbf{R}_\mathcal{R}$, respectively (i.e. potential mutual confounding among all variables internal to $\mathcal{W}_\mathcal{L}$ and $\mathcal{W}_\mathcal{R}$, respectively). 
We have, $\forall\mathbf{r}\in\nu(\mathbf{R})$, $\forall\mathbf{w}\in\nu(\mathbf{W})$ (so in particular, $\P\{\mathbf{R}_\mathcal{L}=\mathbf{r}_\mathcal{L}\},\P\{\mathbf{R}_\mathcal{R}=\mathbf{r}_\mathcal{R}\},\P\{\mathbf{W}_\mathcal{L}=\mathbf{w}_\mathcal{L}\},\P\{\mathbf{W}_\mathcal{L}=\mathbf{w}_\mathcal{L},\mathbf{R}_\mathcal{R}=\mathbf{r}_\mathcal{R}\}=\P\{\mathbf{W}_\mathcal{L}=\mathbf{w}_\mathcal{L}\}\P\{\mathbf{R}_\mathcal{R}=\mathbf{r}_\mathcal{R}\}>0$), 
\begin{align*}
    \P\{\mathbf{W}=\mathbf{w},\mathbf{R}=\mathbf{r}\}
    &=\P\{\mathbf{W}_\mathcal{L}=\mathbf{w}_\mathcal{L},\mathbf{W}_\mathcal{R}=\mathbf{w}_\mathcal{R},\mathbf{R}_\mathcal{L}=\mathbf{r}_\mathcal{L},\mathbf{R}_\mathcal{R}=\mathbf{r}_\mathcal{R}\}\\
    &=\P\{\mathbf{W}_\mathcal{L}=\mathbf{w}_\mathcal{L}\mid \mathbf{R}_\mathcal{L}=\mathbf{r}_\mathcal{L}\}\P\{\mathbf{R}_\mathcal{L}=\mathbf{r}_\mathcal{L}\}\\
    &\quad\quad\P\{\mathbf{W}_\mathcal{R}=\mathbf{w}_\mathcal{R}\mid \mathbf{W}_\mathcal{L}=\mathbf{w}_\mathcal{L},\mathbf{R}_\mathcal{R}=\mathbf{r}_\mathcal{R}\}\P\{\mathbf{R}_\mathcal{R}=\mathbf{r}_\mathcal{R}\}.\\
\end{align*}
So $\forall\mathbf{w}\in\nu(\mathbf{W})$, 
\begin{align*}
    \P\{\mathbf{W}=\mathbf{w}\}
    &=\sum_{\mathbf{r}\in\nu(\mathbf{R})}\P\{\mathbf{W}=\mathbf{w},\mathbf{R}=\mathbf{r}\}\\
    &=\sum_{\mathbf{r}\in\nu(\mathbf{R})}\P\{\mathbf{W}_\mathcal{L}=\mathbf{w}_\mathcal{L}\mid \mathbf{R}_\mathcal{L}=\mathbf{r}_\mathcal{L}\}\P\{\mathbf{R}_\mathcal{L}=\mathbf{r}_\mathcal{L}\}\\
    &\quad\quad\quad\quad\quad\P\{\mathbf{W}_\mathcal{R}=\mathbf{w}_\mathcal{R}\mid \mathbf{W}_\mathcal{L}=\mathbf{w}_\mathcal{L},\mathbf{R}_\mathcal{R}=\mathbf{r}_\mathcal{R}\}\P\{\mathbf{R}_\mathcal{R}=\mathbf{r}_\mathcal{R}\}\\
    &=\sum_{\mathbf{r}_\mathcal{L}\in\nu(\mathbf{R}_\mathcal{L})}\sum_{\mathbf{r}_\mathcal{R}\in\nu(\mathbf{R}_\mathcal{R})}\P\{\mathbf{W}_\mathcal{L}=\mathbf{w}_\mathcal{L}\mid \mathbf{R}_\mathcal{L}=\mathbf{r}_\mathcal{L}\}\P\{\mathbf{R}_\mathcal{L}=\mathbf{r}_\mathcal{L}\}\\
    &\quad\quad\quad\quad\quad\quad\quad\quad\quad\P\{\mathbf{W}_\mathcal{R}=\mathbf{w}_\mathcal{R}\mid \mathbf{W}_\mathcal{L}=\mathbf{w}_\mathcal{L},\mathbf{R}_\mathcal{R}=\mathbf{r}_\mathcal{R}\}\P\{\mathbf{R}_\mathcal{R}=\mathbf{r}_\mathcal{R}\}\\
    &=\sum_{\mathbf{r}_\mathcal{L}\in\nu(\mathbf{R}_\mathcal{L})}\P\{\mathbf{W}_\mathcal{L}=\mathbf{w}_\mathcal{L}\mid \mathbf{R}_\mathcal{L}=\mathbf{r}_\mathcal{L}\}\P\{\mathbf{R}_\mathcal{L}=\mathbf{r}_\mathcal{L}\}\\
    &\quad\quad\sum_{\mathbf{r}_\mathcal{R}\in\nu(\mathbf{R}_\mathcal{R})}\P\{\mathbf{W}_\mathcal{R}=\mathbf{w}_\mathcal{R}\mid \mathbf{W}_\mathcal{L}=\mathbf{w}_\mathcal{L},\mathbf{R}_\mathcal{R}=\mathbf{r}_\mathcal{R}\}\P\{\mathbf{R}_\mathcal{R}=\mathbf{r}_\mathcal{R}\}\\
    &=\P\{\mathbf{W}_\mathcal{L}=\mathbf{w}_\mathcal{L}\}\sum_{\mathbf{r}_\mathcal{R}\in\nu(\mathbf{R}_\mathcal{R})}\P\{\mathbf{W}_\mathcal{R}=\mathbf{w}_\mathcal{R}\mid \mathbf{W}_\mathcal{L}=\mathbf{w}_\mathcal{L},\mathbf{R}_\mathcal{R}=\mathbf{r}_\mathcal{R}\}\P\{\mathbf{R}_\mathcal{R}=\mathbf{r}_\mathcal{R}\}.\\
\end{align*}

Hence, $\forall b\in\{1,\dots,B\}$, 
\begin{align*}
    p_b
    &=P\{\mathbf{W}_\mathcal{R}=\mathbf{w}_{b,\mathcal{R}}\mid \mathbf{W}_\mathcal{L}=\mathbf{w}_{b,\mathcal{L}}\}\\
    &=\frac{\P\{\mathbf{W}_\mathcal{L}=\mathbf{w}_{b,\mathcal{L}},\mathbf{W}_\mathcal{R}=\mathbf{w}_{b,\mathcal{R}}\}}{\P\{\mathbf{W}_\mathcal{L}=\mathbf{w}_{b,\mathcal{L}}\}}\\
    &=\frac{\P\{\mathbf{W}=\mathbf{w}_b\}}{\P\{\mathbf{W}_\mathcal{L}=\mathbf{w}_{b,\mathcal{L}}\}}\\
    &=\frac{\P\{\mathbf{W}_\mathcal{L}=\mathbf{w}_{b,\mathcal{L}}\}\sum_{\gamma=1}^{\aleph_\mathcal{R}}\P\{\mathbf{W}_\mathcal{R}=\mathbf{w}_{b,\mathcal{R}}\mid \mathbf{W}_\mathcal{L}=\mathbf{w}_{b,\mathcal{L}},\mathbf{R}_\mathcal{R}=\mathbf{r}_\gamma\}\P\{\mathbf{R}_\mathcal{R}=\mathbf{r}_\gamma\}}{\P\{\mathbf{W}_\mathcal{L}=\mathbf{w}_{b,\mathcal{L}}\}}\\
    &=\sum_{\gamma=1}^{\aleph_\mathcal{R}}\P\{\mathbf{W}_\mathcal{R}=\mathbf{w}_{b,\mathcal{R}}\mid \mathbf{W}_\mathcal{L}=\mathbf{w}_{b,\mathcal{L}},\mathbf{R}_\mathcal{R}=\mathbf{r}_\gamma\}\P\{\mathbf{R}_\mathcal{R}=\mathbf{r}_\gamma\}\\
    &=\sum_{\gamma=1}^{\aleph_\mathcal{R}}P_{b\gamma}q_\gamma\\
\end{align*}
 where $P\in\{0,1\}^{B\times\aleph_\mathcal{R}}$ is given by $\forall b\in\{1,\dots,B\},\gamma\in\{1,\dots,\aleph_\mathcal{R}\}$, 
\begin{align*}
    P_{b\gamma}
    &:=\P\{\mathbf{W}_\mathcal{R}=\mathbf{w}_{b,\mathcal{R}}\mid \mathbf{W}_\mathcal{L}=\mathbf{w}_{b,\mathcal{L}},\mathbf{R}_\mathcal{R}=\mathbf{r}_\gamma\}\\
    &=\begin{cases}1&\text{if }\forall i\in\mathcal{R},w_i=g_{W_i}(\mathbf{w}_{b,\mathcal{L}},\mathbf{r}_\gamma)\\0&\text{otherwise}\end{cases}.\\
\end{align*}

Moreover, $\forall b\in\{1,\dots,B\}$, 
\begin{align*}
    p^*_b
    &=\P\{\mathbf{W}=\mathbf{w}_b\}\\
    &=\P\{\mathbf{W}_\mathcal{L}=\mathbf{w}_{b,\mathcal{L}},\mathbf{W}_\mathcal{R}=\mathbf{w}_{b,\mathcal{R}}\}\\
    &=\P\{\mathbf{W}_\mathcal{L}=\mathbf{w}_{b,\mathcal{L}}\}\P\{\mathbf{W}_\mathcal{R}=\mathbf{w}_{b,\mathcal{R}}\mid \mathbf{W}_\mathcal{L}=\mathbf{w}_{b,\mathcal{L}}\}\\
    &=\P\{\mathbf{W}_\mathcal{L}=\mathbf{w}_{b,\mathcal{L}}\}p_b\\
    &=\P\{\mathbf{W}_\mathcal{L}=\mathbf{w}_{b,\mathcal{L}}\}\sum_{\gamma=1}^{\aleph_\mathcal{R}}P_{b\gamma}q_\gamma\\
    &=P^*_{b\gamma}q_\gamma\\
\end{align*}
 where $P^*\in[0,1]^{B\times\aleph_\mathcal{R}}$ is given by $\forall b\in\{1,\dots,B\},\gamma\in\{1,\dots,\aleph_\mathcal{R}\}$, 
\begin{align*}
    P^*_{b\gamma}
    &:=\P\{\mathbf{W}=\mathbf{w}_{b}\mid\mathbf{R}_\mathcal{R}=\mathbf{r}_\gamma\}\\
    &=\P\{\mathbf{W}_\mathcal{L}=\mathbf{w}_{b,\mathcal{L}},\mathbf{W}_\mathcal{R}=\mathbf{w}_{b,\mathcal{R}}\mid\mathbf{R}_\mathcal{R}=\mathbf{r}_\gamma\}\\
    &=\P\{\mathbf{W}_\mathcal{R}=\mathbf{w}_{b,\mathcal{R}}\mid\mathbf{W}_\mathcal{L}=\mathbf{w}_{b,\mathcal{L}},\mathbf{R}_\mathcal{R}=\mathbf{r}_\gamma\}\P\{\mathbf{W}_\mathcal{L}=\mathbf{w}_{b,\mathcal{L}}\mid\mathbf{R}_\mathcal{R}=\mathbf{r}_\gamma\}\\
    &=\P\{\mathbf{W}_\mathcal{L}=\mathbf{w}_{b,\mathcal{L}}\}\P\{\mathbf{W}_\mathcal{R}=\mathbf{w}_{b,\mathcal{R}}\mid \mathbf{W}_\mathcal{L}=\mathbf{w}_{b,\mathcal{L}},\mathbf{R}_\mathcal{R}=\mathbf{r}_\gamma\}\\
    &=\P\{\mathbf{W}_\mathcal{L}=\mathbf{w}_{b,\mathcal{L}}\}P_{b\gamma}\\
    &=\begin{cases}\P\{\mathbf{W}_\mathcal{L}=\mathbf{w}_{b,\mathcal{L}}\}&\text{if }\forall i\in\mathcal{R},w_i=g_{W_i}(\mathbf{w}_{b,\mathcal{L}},\mathbf{r}_\gamma)\\0&\text{otherwise}\end{cases}.\\
\end{align*}

Since $\forall b\in\{1,\dots,B\},\ p^*_b=\P\{\mathbf{W}_\mathcal{L}=\mathbf{w}_{b,\mathcal{L}}\}p_b$, we have $\mathbf{p}^*=\Lambda\mathbf{p}$, where $\Lambda\in[0,1]^{B\times B}$ is given by, $\forall b,c\in\{1,\dots,B\}$, $$\Lambda_{bc}:=\begin{cases}\P\{\mathbf{W}_\mathcal{L}=\mathbf{w}_{b,\mathcal{L}}\}&\text{if }b=c\\0&\text{otherwise}\end{cases}$$

Note that $\forall b\in\{1,\dots,B\},\forall\gamma\in\{1,\dots,\aleph_\mathcal{R}\},\sum_{c=1}^{B}\Lambda_{bc}P_{c\gamma}=\Lambda_{bb}P_{b\gamma}=\P\{\mathbf{W}_\mathcal{L}=\mathbf{w}_{b,\mathcal{L}}\}P_{b\gamma}=P^*_{b\gamma}$, so $\Lambda P=P^*$. Note further that the diagonal entries of $\Lambda$ all are non-zero (since $\forall b\in\{1,\dots,B\},\ \mathbf{w}_{b,\mathcal{L}}\in\nu(\mathbf{W}_\mathcal{L})$), so $\Lambda$ is invertible and hence bijectively maps between the conditional probability vector $\mathbf{p}=P\mathbf{q}\in[0,1]^B$ and the corresponding marginal one $\mathbf{p}^*=P^*\mathbf{q}\in[0,1]^B$. Consequently, $\mathbf{p}=P\mathbf{q}\iff\Lambda\mathbf{p}=\Lambda P\mathbf{q}\iff\mathbf{p}^*=P^*\mathbf{q}$.

Since the distribution of the unmeasured influences $U$, or equivalently the response function variables $R$, is independent of the DAG, the DAG cannot encode any quantitative constraints in the form of relationships between these variables. Thus, the structural equations encoded by the DAG can only imply constraints (ignoring the distinction between the left and right sides, since this can be considered within each of those sets) based on the following types of independence relations: (i) $W_i \indep U$ for some $i$, (ii) $W_i \indep U | W_\mathcal{B}$ for some $i$ and set of observed variables $W_\mathcal{B}$, (iii) $W_i \indep W_j$ for some $i,j$, (iv) $W_i \indep W_j | W_\mathcal{A}$ for some $i, j$ and set of observed variables $W_\mathcal{A}$ or (v) $W_i \indep W_j | U$ for some $i, j$. Cases (i) and (ii) imply that $U$ is not a parent of $W_i$, in violation of Condition \ref{item:cond5} or \ref{item:cond6}. Cases (iii) and (iv) imply that $U$ is either not a parent of $W_i$ or not of $W_j$, again in violation of Condition \ref{item:cond5} or \ref{item:cond6}. Case (v) implies that for $i, j$, we have $\P\{W_i, W_j\} = \sum_{R} \P\{W_i, W_j | R\} \P\{R\} = \sum_{R} \P\{W_i | R\} \P\{W_j | R\} \P\{R\}$ which is still linear in $\mathbf{q}$. 

Now relating this last point to the enumeration of constraints above, note the vector $\mathbf{p}^*$ enumerates all joint probabilities of all observed variables in the DAG. Hence, constraints relating linear combinations of $\mathbf{q}$ to joint, conditional, or marginal probabilities of subsets of $\mathcal{W}$ can be directly obtained as transformations among rows of the existing constraints $\mathbf{p}^* = P^*\mathbf{q}$. The addition of those are clearly redundant. In other words, the matrix $P$ contains complete information about any and all relationships between the observed joint distribution and the joint distribution of the response function variables of the $\mathcal{R}$-side that are possible under our conditions.
By the above, the complete set of constraints on observed probabilities is equivalent to a system that is linear in $\mathbf{q}$. 
\end{proof}

\begin{proof}[Proof of Proposition~\ref{t2}]
\label{proofoft2}

Let again $\mathcal{P}=\{i_1,\dots,i_P\}$ and $\mathcal{O}=\{j_1,\dots,j_O\}$ be respectively the indices of the potential and factual outcomes in $Q$, and $\Gamma(Q)=\{\mathbf{r}\in\nu(\mathbf{R}):w_{i_1}=h_{W_{i_1}}^{A_{i_1}}(\mathbf{r},W_{i_1}),\dots,w_{i_P}=h_{W_{i_P}}^{A_{i_P}}(\mathbf{r},W_{i_P}),w_{j_1}=g_{W_{j_1}}(\mathbf{r}),\dots,w_{j_O}=g_{W_{j_O}}(\mathbf{r})\}$. We have $(\mathbf{R}_\mathcal{L}\indep\mathbf{R}_\mathcal{R})_G$ and, by condition \ref{item:cond3}, $\mathcal{P}\cup\mathcal{O}\subset\mathcal{R}$ and if $\mathcal{L}\ne\varnothing$, $\Gamma(Q)=\nu(\mathbf{R}_\mathcal{L})\times\Gamma_\mathcal{R}(Q)$, where $\Gamma_\mathcal{R}(Q):=\{\mathbf{r}_\mathcal{R}\in\nu(\mathbf{R}_\mathcal{R}):w_{i_1}=h_{W_{i_1}}^{A_{i_1}}(\mathbf{r}_\mathcal{R},W_{i_1}),\dots,w_{i_P}=h_{W_{i_P}}^{A_{i_P}}(\mathbf{r}_\mathcal{R},W_{i_P}),w_{j_1}=g_{W_{j_1}}(\mathbf{r}_\mathcal{R}),\dots,w_{j_O}=g_{W_{j_O}}(\mathbf{r}_\mathcal{R})\}$. Condition \ref{item:cond4} ensures that, if $\mathcal{L}$ is not empty, then all paths from the potential outcomes in $Q$ to any variables in $\mathcal{L}$ must pass through the intervention set, thus negating any influence of $\mathbf{R}_\mathcal{L}$ on any of the variables in $Q$. Hence, if $\mathcal{L}=\varnothing$, then 
\begin{align*}
    Q&=\P\{h_{W_{i_1}}^{A_{i_1}}(\mathbf{R},W_{i_1})=w_{i_1},\dots,h_{W_{i_P}}^{A_{i_P}}(\mathbf{R},W_{i_P})=w_{i_P},g_{W_{j_1}}(\mathbf{R})=w_{j_1},\dots,g_{W_{j_O}}(\mathbf{R})=w_{j_O}\}\\
    &=\sum_{\mathbf{r}\in\Gamma(Q)}\P\{\mathbf{R}=\mathbf{r}\}\\
    &=\sum_{\gamma=1}^{\aleph_\mathcal{R}}\mathbb{I}_{\Gamma(Q)}(\mathbf{r}_\gamma)q_\gamma=\alpha^\top\mathbf{q},
\end{align*}
where $\mathbb{I}(\cdot)$ is the indicator function and  $\alpha\in\{0,1\}^{\aleph_\mathcal{R}}$ is given by $$\forall\gamma\in\{1,\dots,\aleph_\mathcal{R}\}, \alpha_\gamma:=\begin{cases}1&\text{if }\mathbf{r}_\gamma\in\Gamma(Q)\\0&\text{otherwise}\end{cases}.$$

If $\mathcal{L}\ne\varnothing$, we have 
\begin{align*}
    Q&=\P\{h_{W_{i_1}}^{A_{i_1}}(\mathbf{R},W_{i_1})=w_{i_1},\dots,h_{W_{i_P}}^{A_{i_P}}(\mathbf{R},W_{i_P})=w_{i_P}\}\\
    &=\sum_{\mathbf{r}\in\Gamma(Q)}\P\{\mathbf{R}=\mathbf{r}\}\\
    &=\sum_{\mathbf{r}\in\Gamma(Q)}\P\{\mathbf{R}_\mathcal{L}=\mathbf{r}_\mathcal{L},\mathbf{R}_\mathcal{R}=\mathbf{r}_\mathcal{R}\}\\
    &=\sum_{\mathbf{r}\in\Gamma(Q)}\P\{\mathbf{R}_\mathcal{L}=\mathbf{r}_\mathcal{L}\}\P\{\mathbf{R}_\mathcal{R}=\mathbf{r}_\mathcal{R}\}\\
    &=\sum_{(\mathbf{r}_\mathcal{L},\mathbf{r}_\mathcal{R})\in\nu(\mathbf{R}_\mathcal{L})\times\Gamma_\mathcal{R}(Q)}\P\{\mathbf{R}_\mathcal{L}=\mathbf{r}_\mathcal{L}\}\P\{\mathbf{R}_\mathcal{R}=\mathbf{r}_\mathcal{R}\}\\
    &=\sum_{\mathbf{r}_\mathcal{L}\in\nu(\mathbf{R}_\mathcal{L})}\sum_{\mathbf{r}_\mathcal{R}\in\Gamma_\mathcal{R}(Q)}\P\{\mathbf{R}_\mathcal{L}=\mathbf{r}_\mathcal{L}\}\P\{\mathbf{R}_\mathcal{R}=\mathbf{r}_\mathcal{R}\}\\
    &=\sum_{\mathbf{r}_\mathcal{L}\in\nu(\mathbf{R}_\mathcal{L})}\P\{\mathbf{R}_\mathcal{L}=\mathbf{r}_\mathcal{L}\}\sum_{\mathbf{r}_\mathcal{R}\in\Gamma_\mathcal{R}(Q)}\P\{\mathbf{R}_\mathcal{R}=\mathbf{r}_\mathcal{R}\}\\
    &=\sum_{\mathbf{r}_\mathcal{R}\in\Gamma_\mathcal{R}(Q)}\P\{\mathbf{R}_\mathcal{R}=\mathbf{r}_\mathcal{R}\}\\
    &=\sum_{\gamma=1}^{\aleph_\mathcal{R}}\mathbb{I}_{\Gamma_\mathcal{R}(Q)}(\mathbf{r}_\gamma)q_\gamma=\alpha^\top\mathbf{q},
\end{align*}
where $\alpha\in\{0,1\}^{\aleph_\mathcal{R}}$ is given by $\forall\gamma\in\{1,\dots,\aleph_\mathcal{R}\}$, $\alpha_\gamma:=\begin{cases}1&\text{if }\mathbf{r}_\gamma\in\Gamma_\mathcal{R}(Q)\\0&\text{otherwise}\end{cases}$.

%By Assumption $(5)$, all of the variables occurring in $Q$ are in $\mathcal{W}_\mathcal{R}$. Thus by Assumptions $(1)$ and $(4)$, these variables are conditionally independent of the response function variables in $\mathbf{R}_\mathcal{L}$ given $W_\mathcal{L}$. However, by Assumption $(6.2)$, the intervention set $d$-separates $W_\mathcal{R}$ from $\mathbf{R}_\mathcal{L}$. Also, by Assumption $(4)$, $\mathbf{R}_\mathcal{L}$ is independent of $\mathbf{R}_\mathcal{R}$. 

\end{proof}

\begin{proof}[Proof of Proposition \ref{tightness}]
\label{proofoftightness}

Proposition \ref{linprobth} ensures that the linear constraints $\mathbf{p}^* = P^*\mathbf{q}$ are necessary and sufficient for the probability distribution to be compatible with the causal model. Solving the optimization problem with these constraints is equivalent to solving it with the constraints $\mathbf{p} = P\mathbf{q}$ because the relation is obtained by multiplying both sides of the equation by an invertible constant matrix. Proposition \ref{t2} demonstrates that the objective function is linear in $\mathbf{q}$. The constraint space is closed and non-empty, and is bounded by the probabilistic constraints. Subject to any additional linear constraints specified in the form of equalities or non-strict inequalities, the constraint space is closed and bounded, hence compact, so by the extreme value theorem and the fact that the objective is linear, hence continuous, the primal problem has an optimal feasible solution. By the strong duality theorem, the dual problem has a global optimum coinciding with that of the primal, and again has a bounded constraint space, so by the fundamental theorem of linear programming, it can be found in terms of $\mathbf{p}$ via vertex enumeration.  %In order to show that they are tight, we must argue that under the Assumptions (\ref{item:cond5}) and (\ref{item:cond6}), the constraints are sufficient, i.e., the feasible space is not subject to any further constraints. 

%Assumptions (\ref{item:cond5}) and (\ref{item:cond6}) ensure that the components of $\mathbf{q}$  which are probabilities of the form $\P\{\mathbf{R}_\mathcal{R} = \mathbf{r}_\gamma\}$ cannot be further factorized into products of distinct parameters such as $\P\{\mathbf{R}_{\mathcal{A}} = \mathbf{r}_\mathcal{A}\}\P\{\mathbf{R}_\mathcal{B} = \mathbf{r}_\mathcal{B}\}$ for some partition $(\mathcal{A}, \mathcal{B})$ of $\mathcal{R}$. Such as factorization would lead to an objective and/or constraints that are nonlinear in the unknown parameters. Since that is not the case under these assumptions,

\end{proof}

\end{document}

\jname{Biometrika}
%% The year, volume, and number are determined on publication
\jyear{2020}
\jvol{103}
\jnum{1}
%% The \doi{...} and \accessdate commands are used by the production team
%\doi{10.1093/biomet/asm023}
\accessdate{Advance Access publication on 31 July 2018}

%% These dates are usually set by the production team
\received{8 April 2020}
\revised{1 April 2021}

%% The left and right page headers are defined here:
\markboth{M.C. Sachs et~al.}{Symbolic Computation of Tight Causal Bounds: Supplemental Appendix}

%% Here are the title, author names and addresses
\title{Symbolic Computation of Tight Causal Bounds: Supplemental Appendix}

\author{M.C. SACHS}
\affil{Department of Medical Epidemiology and Biostatistics \\ Karolinska Institutet \\
Box 281, 17177 Stockholm, Sweden. \email{michael.sachs@ki.se}}

\author{E.E. GABRIEL}
\affil{Department of Medical Epidemiology and Biostatistics \\ Karolinska Institutet \\
Box 281, 17177 Stockholm, Sweden. \email{erin.gabriel@ki.se}}

\author{A. SJÖLANDER}
\affil{Department of Medical Epidemiology and Biostatistics \\ Karolinska Institutet \\
Box 281, 17177 Stockholm, Sweden. \email{arvid.sjolander@ki.se}}

\maketitle

\section{Bounds for the two instrumental variable problem}

The bounds are reported in terms of variables pab\_cd, which represents the probability $\mbox{pr}(X = a, Y = b | Z1 = c, Z2 = d)$. The min and max are taken over the terms that each appear on a separate line. 

\subsection{Lower bound}

{\scriptsize
\begin{verbatim}
max {
p00_00 + p00_01 + p10_01 + p01_01 + 2 p11_01 - 2
p00_00 - p00_10 - p00_01 - p10_10 - p10_01 - p01_10 - p01_01 - p11_01 + 1
p00_00 - p00_11 - p10_11 - p01_11
p00_10 + p00_01 + p10_01 + p01_01 + 2 p11_01 - 2
- p00_10 - p10_10 - p10_01 - p01_10 - p01_01 - p11_01 + 1
- p00_10 + 4 p00_01 + p00_11 - p10_10 + 2 p10_01 + p10_11 + 2 p01_01 + 4 p11_01 - 4
- 2 p00_01 + p00_11 - 2 p10_10 - 2 p10_01 + p10_11 - 2 p01_10 - p01_01 - p11_01 + 1
- p00_10 - 2 p00_01 + p00_11 - 2 p10_10 - 2 p10_01 + p10_11 - 2 p01_10 - 2 p01_01 - 2 p11_01 + 2
3 p00_01 + p00_11 + 2 p10_01 + p10_11 + 2 p01_01 + 4 p11_01 - 4
p00_10 - p00_11 - p10_11 - p01_11
p00_01 - p00_11 - p10_11 - p01_11
p00_00 - p00_10 - 2 p00_01 + p10_00 - 2 p10_10 - 2 p10_01 - 2 p01_10 - 2 p01_01 - 2 p11_01 + 2
p00_00 + 3 p00_01 + p10_00 + 2 p10_01 + 2 p01_01 + 4 p11_01 - 4
p00_00 - p00_10 + 4 p00_01 + p10_00 - p10_10 + 2 p10_01 + 2 p01_01 + 4 p11_01 - 4
p00_00 - 2 p00_01 + p10_00 - 2 p10_10 - 2 p10_01 - 2 p01_10 - p01_01 - p11_01 + 1
- p00_11 - p10_00 - p10_11 - p01_00 - p01_01 - p01_11 - p11_01 + 1
p00_10 - p00_11 - p10_00 + p10_10 - p10_11 - p01_00 - p01_11
- p00_00 - 2 p10_00 - 2 p01_00 - p01_01 - p11_01 + 1
- p00_00 + p00_10 - 2 p10_00 + p10_10 - 2 p01_00
p00_01 + p00_11 + p10_01 + p01_01 + 2 p11_01 - 2
- p00_10 - p00_01 + p00_11 - p10_10 - p10_01 - p01_10 - p01_01 - p11_01 + 1
- p00_00 + p00_01 - p10_00 - p01_00
- p00_00 + p00_10 - p10_00 - p01_00
- p00_01 - p10_10 - p10_01 - p01_10 - p01_01 - p11_01 + 1
2 p00_01 + p10_01 + p01_01 + 2 p11_01 - 2
- p00_00 + 4 p00_01 + p00_11 - p10_00 + 2 p10_01 + p10_11 + 2 p01_01 + 4 p11_01 - 4
- p00_00 - 2 p00_01 + p00_11 - p10_00 - 2 p10_10 - 2 p10_01 + p10_11 - 2 p01_10 - 2 p01_01 - 2 p11_01 + 2
- p00_00 - p10_00 - p10_11 - p01_00 - p01_01 - p01_11 - p11_01 + 1
- p00_00 + p00_10 - p10_00 + p10_10 - p10_11 - p01_00 - p01_11
- p00_00 + p00_11 - p10_00 - p01_00
- p00_11 - 2 p10_00 - p10_11 - 2 p01_00 - p01_01 - p11_01 + 1
p00_10 - p00_11 - 2 p10_00 + p10_10 - p10_11 - 2 p01_00
p00_00 - p00_11 + p10_00 - 2 p10_11 - 2 p01_11
- p00_00 - p10_00 - 2 p10_11 - p01_01 - 2 p01_11 - p11_01 + 1
- p00_00 + p00_10 - p10_00 + p10_10 - 2 p10_11 - 2 p01_11
p00_01 + p00_11 - p10_00 + p10_01 + p10_11 - p01_00 + p01_01 + 2 p11_01 - 2
- p00_10 - p00_01 + p00_11 - p10_00 - p10_10 - p10_01 + p10_11 - p01_00 - p01_10 - p01_01 - p11_01 + 1
- p00_00 + p00_11 - 2 p10_00 + p10_11 - 2 p01_00
- p00_10 + p00_11 - 2 p10_00 - p10_10 + p10_11 - 2 p01_00
p00_11 - 2 p10_00 + p10_11 - 2 p01_00 + p01_01 + p11_01 - 1
- p00_10 - 2 p10_00 - p10_10 - 2 p01_00 - p01_01 - p11_01 + 1
p00_10 - 2 p10_00 + p10_10 - 2 p01_00 + p01_01 + p11_01 - 1
- p10_00 - p01_00
p00_00 + p10_00 - p10_10 - p01_10 + p11_01 - 1
p00_00 - p00_10 + p00_01 + p10_00 - p10_10 - p01_10 + p11_01 - 1
p00_00 - p00_11 - p10_00 - p10_11 - p01_00
p00_00 - p00_10 - p10_00 - p10_10 - p01_00
p00_00 - p10_00 - p01_00 + p01_01 + p11_01 - 1
- p00_00 + 2 p00_01 + p00_11 - p10_00 + p10_01 + p10_11 - p01_00 + p01_01 + 2 p11_01 - 2
- p00_00 - p00_01 + p00_11 - p10_00 - p10_10 - p10_01 + p10_11 - p01_00 - p01_10 - p01_01 - p11_01 + 1
- p00_00 + p00_10 - p00_01 - p10_00 - p10_10 - p10_01 - p01_10 - p01_01 - p11_01 + 1
- p00_00 + 3 p00_01 - p10_00 + p10_01 + p01_01 + 2 p11_01 - 2
- p00_10 - 2 p00_01 - 2 p10_10 - 2 p10_01 - 2 p01_10 - 3 p01_01 - 3 p11_01 + 3
p00_10 + 3 p00_01 + p10_10 + 2 p10_01 + 2 p01_01 + 4 p11_01 - 4
- p00_00 + p00_10 + 2 p00_01 - p10_00 + p10_01 + p01_01 + 2 p11_01 - 2
- p00_00 - p10_00 - p10_10 - p10_01 - p01_10 - p01_01 - p11_01 + 1
- p00_00 - p00_01 + p00_11 - p10_00 - p10_10 - p10_01 - p01_10 - p01_01 - p11_01 + 1
- p00_00 + 2 p00_01 + p00_11 - p10_00 + p10_01 + p01_01 + 2 p11_01 - 2
- p00_00 - 2 p00_01 - p10_00 - 2 p10_10 - 2 p10_01 - 2 p01_10 - 3 p01_01 - 3 p11_01 + 3
- p00_00 + p00_10 + 4 p00_01 - p10_00 + p10_10 + 2 p10_01 + 2 p01_01 + 4 p11_01 - 4
- p00_10 + p00_11 - p10_00 - p10_10 - p01_00
p00_11 - p10_00 - p01_00 + p01_01 + p11_01 - 1
- p00_10 + p00_01 - p10_00 - p10_10 - p01_00
p00_10 - p10_00 - p01_00 + p01_01 + p11_01 - 1
- p00_11 - 2 p10_11 - p01_01 - 2 p01_11 - p11_01 + 1
p00_10 - p00_11 + p10_10 - 2 p10_11 - 2 p01_11
- 2 p00_01 - p00_11 - 2 p10_10 - 2 p10_01 - p10_11 - 2 p01_10 - 3 p01_01 - 3 p11_01 + 3
p00_10 + 4 p00_01 - p00_11 + p10_10 + 2 p10_01 - p10_11 + 2 p01_01 + 4 p11_01 - 4
p00_11 - p10_10 + p10_11 - p01_10 + p11_01 - 1
- p00_10 + p00_01 + p00_11 - p10_10 + p10_11 - p01_10 + p11_01 - 1
- p00_10 - p00_01 - p10_00 - p10_10 - p10_01 - p01_00 - p01_10 - 2 p01_01 - 2 p11_01 + 2
p00_10 + p00_01 - p10_00 + p10_10 + p10_01 - p01_00 + p01_01 + 2 p11_01 - 2
p00_10 - p00_01 - p10_10 - p10_01 - p01_10
- p00_10 + 3 p00_01 - p10_10 + p10_01 + p01_01 + 2 p11_01 - 2
- p00_00 + p00_11 - p10_00 - p10_11 - p01_11
p00_00 - p00_01 - p10_10 - p10_01 - p01_10
p00_00 - p00_10 + 2 p00_01 - p10_10 + p10_01 + p01_01 + 2 p11_01 - 2
- p00_01 + p00_11 - p10_10 - p10_01 - p01_10
- p00_10 + 2 p00_01 + p00_11 - p10_10 + p10_01 + p01_01 + 2 p11_01 - 2
p00_10 - p00_11 - p10_00 - p10_11 - p01_00
p00_01 - p00_11 - p10_00 - p10_11 - p01_00
- p00_10 - p10_10 - 2 p10_11 - p01_01 - 2 p01_11 - p11_01 + 1
p00_10 + p10_10 - 2 p10_11 + p01_01 - 2 p01_11 + p11_01 - 1
- p00_00 + p00_01 - p10_00 - p10_11 - p01_11
- p00_00 + p00_10 - p10_00 - p10_11 - p01_11
- p00_01 - p00_11 - p10_10 - p10_01 - p10_11 - p01_10 - 2 p01_01 - p01_11 - 2 p11_01 + 2
p00_10 + 2 p00_01 - p00_11 + p10_10 + p10_01 - p10_11 + p01_01 - p01_11 + 2 p11_01 - 2
p00_00 - 2 p00_01 - p00_11 + p10_00 - 2 p10_10 - 2 p10_01 - p10_11 - 2 p01_10 - 2 p01_01 - 2 p11_01 + 2
p00_00 + 4 p00_01 - p00_11 + p10_00 + 2 p10_01 - p10_11 + 2 p01_01 + 4 p11_01 - 4
p00_00 - p00_10 - p00_01 + p10_00 - p10_10 - p10_01 - p10_11 - p01_10 - p01_01 - p01_11 - p11_01 + 1
p00_00 + p00_01 + p10_00 + p10_01 - p10_11 + p01_01 - p01_11 + 2 p11_01 - 2
p00_00 + 2 p00_01 - p00_11 + p10_00 + p10_01 - p10_11 + p01_01 - p01_11 + 2 p11_01 - 2
p00_00 - p00_01 - p00_11 + p10_00 - p10_10 - p10_01 - p10_11 - p01_10 - p01_01 - p01_11 - p11_01 + 1
- p10_11 - p01_11
- p00_00 - p00_01 - p10_00 - p10_10 - p10_01 - p01_00 - p01_10 - 2 p01_01 - 2 p11_01 + 2
- p00_00 + p00_10 + 2 p00_01 - p10_00 + p10_10 + p10_01 - p01_00 + p01_01 + 2 p11_01 - 2
p00_00 - p00_10 + p10_00 - p10_10 - 2 p10_11 - 2 p01_11
p00_00 + p10_00 - 2 p10_11 + p01_01 - 2 p01_11 + p11_01 - 1
- p00_10 + p00_11 - p10_10 - p10_11 - p01_11
p00_11 - p10_11 + p01_01 - p01_11 + p11_01 - 1
p00_00 - p00_10 - p10_10 - p10_11 - p01_11
p00_00 - p10_11 + p01_01 - p01_11 + p11_01 - 1
- p00_10 + p00_01 - p10_10 - p10_11 - p01_11
p00_10 - p10_11 + p01_01 - p01_11 + p11_01 - 1
- p00_10 - p00_01 - p10_10 - p10_01 - p10_11 - p01_10 - 2 p01_01 - p01_11 - 2 p11_01 + 2
p00_10 + p00_01 + p10_10 + p10_01 - p10_11 + p01_01 - p01_11 + 2 p11_01 - 2
p00_10 - p00_01 - p00_11 - p10_10 - p10_01 - p10_11 - p01_10 - p01_01 - p11_01 + 1
3 p00_01 - p00_11 + p10_01 - p10_11 + p01_01 + 2 p11_01 - 2
p00_10 + 2 p00_01 - p00_11 + p10_01 - p10_11 + p01_01 + 2 p11_01 - 2
- p00_11 - p10_10 - p10_01 - p10_11 - p01_10 - p01_01 - p11_01 + 1
p00_00 - p00_01 - p00_11 - p10_10 - p10_01 - p10_11 - p01_10 - p01_01 - p11_01 + 1
p00_00 + 2 p00_01 - p00_11 + p10_01 - p10_11 + p01_01 + 2 p11_01 - 2
}
\end{verbatim}
}
\subsection{Upper bound}

{\scriptsize
\begin{verbatim}
min {
p00_01 - p10_00 + p10_01 + p11_01
- p00_01 - p10_00 - p10_01 - p01_10 - p01_01 - p11_01 + 2
- p10_00 - p01_00 + 1
- p00_11 - p10_00 - p10_11 - 2 p01_00 + 2
- p10_11 - p01_00 + 1
- p10_10 - p01_00 + 1
p00_01 - p01_00 + p01_01 + p11_01
p00_01 - p10_10 + p10_01 + p11_01
- p10_01 - p01_10 + 1
- p00_10 - p10_00 - p10_10 - 2 p01_00 + 2
- p00_10 + p00_01 - p10_00 - p10_10 + p10_01 - p01_00 + p11_01 + 1
- p10_00 - 2 p01_00 + p01_01 + p11_01 + 1
- p00_01 - p10_00 - p10_01 - p01_00 - p01_10 + 2
- p00_00 + 3 p00_01 - p10_00 + 2 p10_01 + p01_01 + 3 p11_01 - 1
- p00_00 - 2 p00_01 - p10_00 - p10_10 - 2 p10_01 - 2 p01_10 - 2 p01_01 - 2 p11_01 + 4
- p00_10 + 3 p00_01 - p10_10 + 2 p10_01 + p01_01 + 3 p11_01 - 1
- 2 p00_01 - p10_10 - 2 p10_01 - 2 p01_10 - p01_01 - p11_01 + 3
- 2 p10_00 - p01_00 - p01_01 - p11_01 + 2
p00_10 - 2 p10_00 + p10_10 - p01_00 + 1
- p00_11 - 2 p10_00 - p10_11 - 2 p01_00 - p01_01 - p11_01 + 3
p00_10 - p00_11 - 2 p10_00 + p10_10 - p10_11 - 2 p01_00 + 2
- p00_10 - 2 p10_00 - p10_10 - 2 p01_00 - p01_01 - p11_01 + 3
p00_10 - 2 p10_00 + p10_10 - 2 p01_00 + p01_01 + p11_01 + 1
- p00_10 + p00_11 - 2 p10_00 - p10_10 + p10_11 - 2 p01_00 + 2
p00_11 - 2 p10_00 + p10_11 - 2 p01_00 + p01_01 + p11_01 + 1
p00_01 - p00_11 - p10_00 + p10_01 - p10_11 - p01_00 + p11_01 + 1
- p00_01 - p00_11 - p10_00 - p10_01 - p10_11 - p01_00 - p01_10 - p01_01 - p11_01 + 3
- 2 p00_01 - p00_11 - 2 p10_10 - 2 p10_01 - p10_11 - 2 p01_10 - 3 p01_01 - 3 p11_01 + 5
p00_10 + 4 p00_01 - p00_11 + p10_10 + 2 p10_01 - p10_11 + 2 p01_01 + 4 p11_01 - 2
p00_00 + 3 p00_01 + p10_00 + p10_01 + 2 p01_01 + 3 p11_01 - 2
p00_00 - p00_01 + p10_00 - 2 p10_10 - p10_01 - p01_10 - p01_01 - p11_01 + 2
- p00_01 - p10_01 - p10_11 - p01_10 - p01_01 - p11_01 + 2
p00_01 + p10_01 - p10_11 + p11_01
- 2 p00_01 - p00_11 - p10_10 - 2 p10_01 - p10_11 - 2 p01_10 - 2 p01_01 - 2 p11_01 + 4
3 p00_01 - p00_11 + 2 p10_01 - p10_11 + p01_01 + 3 p11_01 - 1
- p00_10 + 2 p00_01 - p10_10 + p10_01 - p01_00 + p01_01 + 2 p11_01
- p00_01 - p10_10 - p10_01 - p01_00 - p01_10 + 2
p00_00 + 4 p00_01 - p00_11 + p10_00 + 2 p10_01 - p10_11 + 2 p01_01 + 4 p11_01 - 2
p00_00 - 2 p00_01 - p00_11 + p10_00 - 2 p10_10 - 2 p10_01 - p10_11 - 2 p01_10 - 2 p01_01 - 2 p11_01 + 4
p00_00 - p00_10 + 4 p00_01 + p10_00 - p10_10 + 2 p10_01 + 2 p01_01 + 4 p11_01 - 2
p00_00 - 2 p00_01 + p10_00 - 2 p10_10 - 2 p10_01 - 2 p01_10 - p01_01 - p11_01 + 3
- p00_00 - 2 p00_01 - p10_00 - 2 p10_10 - 2 p10_01 - 2 p01_10 - 3 p01_01 - 3 p11_01 + 5
- p00_00 + p00_10 + 4 p00_01 - p10_00 + p10_10 + 2 p10_01 + 2 p01_01 + 4 p11_01 - 2
- p10_00 - p10_11 - p01_00 - p01_01 - p11_01 + 2
p00_10 - p10_00 + p10_10 - p10_11 - p01_00 + 1
- p10_00 - p10_10 - p01_00 - p01_01 - p11_01 + 2
p00_10 + p00_01 - p10_00 + p10_10 - p01_00 + p01_01 + p11_01
- p00_01 - 2 p10_10 - p10_01 - p01_10 - 2 p01_01 - 2 p11_01 + 3
p00_10 + 3 p00_01 + p10_10 + p10_01 + 2 p01_01 + 3 p11_01 - 2
p00_00 + 2 p00_01 + p10_00 - p10_10 + p10_01 + p01_01 + 2 p11_01 - 1
p00_00 + p10_00 - p10_10 - p10_01 - p01_10 + 1
p00_00 - p00_01 + p10_00 - p10_10 - p10_01 - p10_11 - p01_10 - p01_01 - p11_01 + 2
p00_00 + 2 p00_01 + p10_00 + p10_01 - p10_11 + p01_01 + 2 p11_01 - 1
- p00_01 - p00_11 - p10_10 - p10_01 - p10_11 - p01_00 - p01_10 - p01_01 - p11_01 + 3
2 p00_01 - p00_11 + p10_01 - p10_11 - p01_00 + p01_01 + 2 p11_01
- p00_00 - 2 p00_01 + p00_11 - p10_00 - 2 p10_10 - 2 p10_01 + p10_11 - 2 p01_10 - 2 p01_01 - 2 p11_01 + 4
- p00_00 + 4 p00_01 + p00_11 - p10_00 + 2 p10_01 + p10_11 + 2 p01_01 + 4 p11_01 - 2
- p00_00 - p10_00 - p10_10 - p01_10 - p01_01 + 2
- p00_00 + p00_01 - p10_00 - p01_10 + p11_01 + 1
- p00_10 + 4 p00_01 + p00_11 - p10_10 + 2 p10_01 + p10_11 + 2 p01_01 + 4 p11_01 - 2
- 2 p00_01 + p00_11 - 2 p10_10 - 2 p10_01 + p10_11 - 2 p01_10 - p01_01 - p11_01 + 3
- p00_01 - p10_10 - p10_01 - p01_10 - p01_01 - p11_01 + 2
2 p00_01 + p10_01 + p01_01 + 2 p11_01 - 1
- p00_01 + p00_11 - 2 p10_10 - p10_01 + p10_11 - p01_10 - p01_01 - p11_01 + 2
3 p00_01 + p00_11 + p10_01 + p10_11 + 2 p01_01 + 3 p11_01 - 2
- p00_10 - p10_10 - p10_11 - 2 p01_11 + 2
- p10_11 + p01_01 - 2 p01_11 + p11_01 + 1
- p10_10 - p01_11 + 1
p00_01 + p01_01 - p01_11 + p11_01
- p00_01 - p10_00 - p10_10 - p10_01 - p01_10 - 2 p01_01 - 2 p11_01 + 3
p00_10 + 2 p00_01 - p10_00 + p10_10 + p10_01 + p01_01 + 2 p11_01 - 1
p00_11 - 2 p10_00 + p10_11 - p01_00 + 1
p00_11 - p10_00 - p10_10 + p10_11 - p01_00 + 1
p00_01 + p00_11 - p10_00 + p10_11 - p01_00 + p01_01 + p11_01
p00_00 - p00_10 + p10_00 - p10_10 - 2 p10_11 - 2 p01_11 + 2
p00_00 + p10_00 - 2 p10_11 + p01_01 - 2 p01_11 + p11_01 + 1
- p00_01 - p10_10 - p10_01 - p10_11 - p01_10 - 2 p01_01 - 2 p11_01 + 3
p00_10 + 2 p00_01 + p10_10 + p10_01 - p10_11 + p01_01 + 2 p11_01 - 1
- 2 p10_11 - p01_01 - p01_11 - p11_01 + 2
p00_10 + p10_10 - 2 p10_11 - p01_11 + 1
2 p00_01 + p00_11 - p10_10 + p10_01 + p10_11 + p01_01 + 2 p11_01 - 1
p00_11 - p10_10 - p10_01 + p10_11 - p01_10 + 1
- p00_00 - p10_00 - 2 p10_11 - p01_01 - 2 p01_11 - p11_01 + 3
- p00_00 + p00_10 - p10_00 + p10_10 - 2 p10_11 - 2 p01_11 + 2
- p00_00 - p00_01 - p10_00 - p10_01 - p10_11 - p01_10 - p01_01 - p01_11 - p11_01 + 3
- p00_00 + p00_01 - p10_00 + p10_01 - p10_11 - p01_11 + p11_01 + 1
- p00_10 - p10_10 - 2 p10_11 - p01_01 - 2 p01_11 - p11_01 + 3
p00_10 + p10_10 - 2 p10_11 + p01_01 - 2 p01_11 + p11_01 + 1
- p10_00 - p10_11 - p01_01 - p01_11 - p11_01 + 2
p00_10 - p10_00 + p10_10 - p10_11 - p01_11 + 1
p00_00 + p10_00 - 2 p10_11 - p01_11 + 1
- p10_00 - p01_11 + 1
- p00_00 - p10_00 - p10_11 - 2 p01_11 + 2
- p00_10 - p10_00 - p10_10 - p01_00 - p01_11 + 2
- p10_00 - p01_00 + p01_01 - p01_11 + p11_01 + 1
- p00_10 + 2 p00_01 - p10_10 + p10_01 + p01_01 - p01_11 + 2 p11_01
- p00_01 - p10_10 - p10_01 - p01_10 - p01_11 + 2
- p00_10 + p00_01 - p10_10 + p10_01 - p10_11 - p01_11 + p11_01 + 1
- p00_01 - p10_01 - p10_11 - p01_10 - p01_11 + 2
- p00_00 - p00_01 - p10_00 - p10_10 - p10_01 - p01_10 - p01_01 - p01_11 - p11_01 + 3
- p00_00 + 2 p00_01 - p10_00 + p10_01 + p01_01 - p01_11 + 2 p11_01
- p00_01 + p00_11 - p10_00 - p10_10 - p10_01 + p10_11 - p01_10 - p01_01 - p11_01 + 2
2 p00_01 + p00_11 - p10_00 + p10_01 + p10_11 + p01_01 + 2 p11_01 - 1
- p00_10 - p10_10 - p10_11 - p01_00 - p01_11 + 2
- p10_11 - p01_00 + p01_01 - p01_11 + p11_01 + 1
- p00_11 - p10_10 - p10_11 - p01_10 - p01_01 + 2
p00_01 - p00_11 - p10_11 - p01_10 + p11_01 + 1
p00_00 + p10_00 - p10_10 - p10_11 - p01_11 + 1
p00_00 + p00_01 + p10_00 - p10_11 + p01_01 - p01_11 + p11_01
- p10_11 - p01_11 + 1
- p10_10 - p10_11 - p01_01 - p01_11 - p11_01 + 2
p00_10 + p00_01 + p10_10 - p10_11 + p01_01 - p01_11 + p11_01
}
\end{verbatim}
}